\documentclass[11pt,showpacs,showkeys,preprintnumbers]{revtex4-2}
\usepackage{graphicx}   % need for figures
\usepackage{color}      % use if color is used in text
\usepackage{braket}
\usepackage{amsmath}
\usepackage{amssymb}
\usepackage{mathrsfs} 
\usepackage{titlesec}
\usepackage{makecell}
\usepackage{natbib,hyperref}
\usepackage[mathscr]{euscript}

\def \be  {\begin{equation}}
\def \ee  {\end{equation}}
\def \ee  {\end{equation}}
\def \bea {\begin{eqnarray}}
\def \eea {\end{eqnarray}}

\newcommand{\de}{\partial}
\newcommand{\R}{\mathbb{R}}
\DeclareMathOperator{\sech}{sech}
\usepackage{mathrsfs}  

\definecolor{ao-english}{rgb}{0.0, 0.5, 0.0}
\definecolor{cadmiumgreen}{rgb}{0.0, 0.42, 0.24}

\newcommand{\nn}{\nonumber}

\begin{document}

\preprint{ECTP-2024-12}
\preprint{WLCAPP-2024-12}
\preprint{FUE-2024-12}
\hspace{0.05cm}

\title{Vacuum Homogeneous and Nonhomogeneous Metrics with Conventional and Quantized Metric Tensor: Singular or Nonsingular Solution}

\author{A. Tawfik$^{1,2}$}     %% a.tawfik@fue.edu.eg
\email{Corresponding author: a.tawfik@fue.edu.eg}
\author{S. G. Elgendi$^3$} %% salah.ali@fsci.bu.edu.eg, salahelgendi@yahoo.com
\author{M. Hanafy$^4$}     %% mahmoud-nasar@sci.bu.edu.eg

\affiliation{$^1$Basic Science Department, Faculty of Engineering, Future University in Egypt, Fifth Settlement, 11835 New Cairo, Egypt}
\affiliation{$^2$Physics Department, Faculty of Science, Islamic University, 42351 Madinah, Saudi Arabia}
\affiliation{$^3$Mathematics Department, Faculty of Science, Islamic University, 42351 Madinah, Saudi Arabia}
\affiliation{$^4$Department of Physics, Faculty of Science, Benha University, 13518 Benha, Egypt}

\begin{abstract}

To investigate whether the Universe underwent a singularity or maintained a nonsingular state, we carry out analytical and numerical analyses of the evolution of the Raychaudhuri equations in vacuum, alongside homogeneous and nonhomogeneous cosmic backgrounds. The results obtained from the Schwarzschild, Friedmann--Lemaître--Robertson--Walker (FLRW), and Einstein--Gilbert--Straus (EGS) metrics  are systematically compared. Analyzing the results from both, conventional and quantized metric tensor, it revealed insights into the nature of initial and spatial singularities. Results associated with the Schwarzschild metric demonstrate a positive evolution that corresponds with a reduction in radial distance (nonsingularity). In contrast, the proposed quantization reverses this trend, leading to a negative evolution (singularity). The situation is similar for the FLRW metric, where the suggested quantization results in a positive evolution as cosmic time decreases, in contrast to the classical and hybrid metrics, which are associated with negative evolution. The analysis of the EGS metric reveals that classical evolution remains positively oriented, particularly with a reduction in radial distance. Moreover, the introduction of quantized and hybrid metric tensors fully retrains the cosmic time dependence. The results obtained are a rightful recognition of the substantial efforts dedicated to the establishment of the Swiss-cheese model, demonstrating that the EGS metric indeed facilitates the temporal and spatial development of our Universe. 
\end{abstract}


\maketitle

%\tableofcontents

%%%%%%%%%%%%%%%%%%%%%%%%%%%%%%%%%%%%%%%%%%%%%%%
\section{Introduction}
\label{sec:intro}

The singularity theorem proposes that the initial singularity is a common characteristic of the homogeneous solutions \cite{Hawking:1973uf,Misner:1973prb,Penrose:1980ge}. On the other hand, it is commonly believed that the initial singularity could potentially be eliminated by incorporating quantum aspects into gravity \cite{Dzierzak:2009ip, Malkiewicz:2009qv}. In addition, there is a conjecture that the cosmological singularity can be eliminated by adopting the "big bounce" model, which is particularly suitable for nearly homogeneous and isotropic solutions \cite{NOVELLO2008127}. Consequently, under extreme conditions, quantum phenomena are projected to be pivotal in shaping the initial epoch of our Universe. The occurrence of the bang or the bounce is anticipated to render the conventional forecasts of general relativity obsolete, as quantum effects take precedence over them. In this context, the term 'big bang' refers to the cosmological singularity, whereas 'big bounce' implies a potential resolution of the cosmological singularity. However, the current investigation does not concentrate on the observational traits of these phases.
 
The current investigation presents a comprehensive examination of two distinct aspects: i) the analysis of vacuum, homogeneous, and nonhomogeneous solutions of the Einstein field equations (EFE), and ii) exploration of the predictions of general relativity with classical and quantum fundamental tensor, $g_{\alpha\beta}$ and $\tilde{g}_{\alpha\beta}$, respectively. The evolution of the Raychaudhuri equations shall be analyzed. Our aim is to compare the results obtained from the Einstein--Gilbert--Straus (EGS) nonhomogeneous metric with the results obtained from the Schwarzschild vacuum and Friedmann--Lemaitre--Robertson--Walker (FLRW) homogeneous metric, respectively. This study is designed to provide a systematic comparison among various categories of EFE solutions, differentiating between those that involve the quantization of the fundamental metric tensor and those that do not.

The solutions to the EFE have a rich historical background, particularly in the matching of spherically symmetric vacuum metric with the FLRW homogeneous and isotropic metric. This matching is done across a spherical boundary, where the Schwarzschild metric is responsible for the interior solution and the FLRW metric for the exterior solution. Noteworthy contributors to this field include Birkhoff and Langer \cite{Birkhoff1923}, McVittie \cite{McVittie:1933zz}, Tolman \cite{Tolman1934}, Einstein and Straus \cite{RevModPhys.17.120}, Bondi \cite{Bondi:1947fta}, and Gilbert \cite{Gilbert1956}. The successful formalism presented by Gilbert  \cite{Gilbert1956,Stewart:1990ngn} is designated as Einstein--Gildert--Straus (EGS) metric \cite{Stuchlik1984}. 

The EGS metric provides a description of the Universe as clusters of cosmic matter distributed in a nonhomogeneous manner and interspersed with voids and holes. Specifically, the cosmic matter is confined within these voids. The quantity of cosmic matter within the voids is equal to the amount excavated to create these voids. Consequently, it is evident that EGS metric offers a broader perspective compared to the conventional cosmological model, enabling the exploration of the effects of nonhomogeneity in the context of an expanding Universe. Similar to the evolution of the nonhomogeneous Universe, which exhibits variations with respect to radial distance $r$ and cosmic time $t$, the timelike geodesic congruence (a family of world-lines) is contingent upon both $r$ and $t$. To explore the interdependence of $r$ and $t$, it is postulated, for simplicity, that the voids housing the entire cosmic substance are spheres, causing the radius to change with the cosmic time. 

The diverse concepts encompassed by the quantized fundamental metric tensor include configurations of duality-symmetry, Finsler--Hamilton manifold, noncommutative algebra, and quantum geometry \cite{Tawfik:2025rel,Tawfik:2025kae,Tawfik:2024itt,NasserTawfik:2024afw,Tawfik:2023onh,Tawfik:2023ugm,Tawfik:2023hdi,Tawfik:2024gtg,Tawfik:2023orl}. The process of incorporating quantum-mechanical revisions begins by generalizing the four-dimensional Riemann manifold to the phase-space structured Finsler--Hamilton manifold. In the Finsler (Hamilton) structure, the auxiliary four-vectors of coordinates and momenta for free-falling quantum particles, $x_0^{\alpha}$ and $p_0^{\beta}$ respectively, are derived from generalized quantum mechanics \cite{Tawfik:2023kxq,Tawfik:2023rrm,Farouk:2023hmi,Tawfik:2021ekh}. The establishment of relativistic gravitization of quantum mechanics in our present study is based on the introduction of the relativistic generalized uncertainty principle (RGUP) \cite{Tawfik:2024gxc,Tawfik:2023kxq}. This leads to the modification of the momentum operator $p_0^{\nu}$ to $\phi(p_0) p_0^{\nu}$, where $\phi(p_0)=1+\beta p_0^{\rho} p_{0 \rho}$ \cite{Tawfik:2023orl,Tawfik:2023kxq}. The RGUP parameter, $\beta$ (not the index), can be determined through table-top experiments and analyzing astronomical observations. Subsequently, the Hessian matrix of the resulting squared Finsler (Hamilton) structure, $F^2(x_0^{\alpha}, \phi(p_0) p_0^{\beta})$, allows for the derivation of the corresponding metric on the Finsler manifold. The resulting metric tensor is eight-dimensional and second-order, incorporating essential quantum aspects. Conversely, on a Riemann manifold, the four-dimensional first-order quantized fundamental tensor can be obtained by matching the line elements of both manifolds. Here, $\rho$ is a dummy index and $\alpha, \beta \in \{0,1,2,3\}$ are free indexes \cite{Tawfik:2023rrm,Tawfik:2023kxq}.

The spacetime curvature is characterized by utilizing the evolution of a set of trajectories \cite{Kolomytsev:1972rf}. The trajectory congruence formed is represented by flow lines that are generated by velocity fields \cite{Tsamparlis:1995nq}. The flow lines represent the trajectories of world lines produced by vector fields, which may exhibit characteristics of being geodesic or nongeodesic \cite{Tawfik:2023onh}.

This manuscript presents novel contributions, particularly in relation to the earlier publications by one of the authors (AT) \cite{Tawfik:2025rel,Tawfik:2025kae,Tawfik:2024itt,NasserTawfik:2024afw,Tawfik:2023onh,Tawfik:2023ugm,Tawfik:2023hdi,Tawfik:2024gtg,Tawfik:2023orl}. This manuscript provides a concise overview of the quantum geometric approach, with further details available in earlier publications. It re-evaluates various homogeneous and nonhomogeneous vacuum solutions to the EFE, utilizing both conventional and quantized metric tensors, while emphasizing the related cosmological and initial singularities \cite{Penrose:1980ge}. This exceptional combination allows for a systematic analysis to disclose the underlying factors contributing to cosmological and initial singularities. The effects of an idealistic versus a realistic configuration of the cosmic background, along with the quantum and classical approximations of the EFE on singularity attenuation, can only be fully understood through this specific study.

The manuscript is organized in the following manner. Section \ref{sec:frmlsm0} presents a comprehensive outline of the formalism, including the introduction of the Schwarzschild, FLRW, and EGS metrics. The solutions to the Einstein Field Equations (EFE) using the conventional metric tensor $g_{\alpha \beta}$ are discussed in Section \ref{sec:WthtQn}. In Section \ref{sec:Raychau2}, the evolution of the Raychaudhuri Equations with the quantized metric tensor $\tilde{g}_{\alpha \beta}$ is derived. Section \ref{sec:dsct} will encompass the numerical results and their discussion. The results concerning the Schwarzschild, FLRW, and EGS metrics with both $g_{\alpha \beta}$ and $\tilde{g}_{\alpha \beta}$ are examined in Sections \ref{sec:Schwrz-dsct}, \ref{sec:RESchFLRW}, and \ref{sec:NrRsltTildaTheta}, respectively. Section \ref{sec:cncl} concludes the manuscript with final thoughts. The appendices present further reviews and derivatives. In Section \ref{sec:tildegmunuApp}, the quantized fundamental metric tensor $\tilde{g}_{\alpha\beta}$ is reviewed. Section \ref{sec:GGA} discusses geometric quantization, geodesic equations, and affine connections. The evolution of Raychaudhuri equations along geodesic congruences is examined in Section \ref{sec:Raychau1}. Section \ref{App:EGA-Dtls} contains detailed numerical results associated with the EGS metric.

\section{Formalism}
\label{sec:frmlsm0}

Various solutions to the Einstein Field Equations have been identified. For a systematic comparison between vacuum, homogeneous and nonhomogeneous cosmic background, Schwarzschild, FLRW, and EGS metrics shall be analyzed with and without geometric quantization. The geometric quantization of general relativity was proposed by one of the authors (AT) \cite{Tawfik:2023ugm,Tawfik:2023hdi,Tawfik:2023kxq,Tawfik:2023rrm,Farouk:2023hmi,Tawfik:2021ekh}. It is based on geometric quantization fundamental metric tensor $\tilde{g}_{\alpha\beta}$, which shall be reviewed in Appendix \ref{sec:tildegmunuApp}. Section \ref{sec:WthtQn} outlines the three metrics with with conventional metric tensor $g_{\alpha \beta}$.

\subsection{Solutions of EFE with Conventional Metric Tensor $g_{\alpha \beta}$}
\label{sec:WthtQn}

\begin{enumerate} 
\item{\bf Schwarzschild Metric}

The Schwarzchild metric is the most general spherically symmetric vacuum exact solution of EFE in a curved spacetime. It characterizes the gravitational field outside a spherical mass $M$ accounting for a complete lack of electric charge and angular momentum and vanishing universal cosmological constant \cite{Birkhoff1923}. In polar spherical coordinates
\begin{equation}
ds^{2}=-\left(1-\frac{2 M}{r}\right) dt^{2}+\frac{dr^{2}}{\left(1-\frac{2 M}{r}\right)}+r^{2}d\Omega^{2}, \label{eq:Schwz1}
\end{equation}
where the metric $d\Omega^2=d\theta^{2}+\sin^{2}{\theta}d\phi^{2}$ characterizes the standard metric on the surface of a two-sphere with a constant radius $r$. 
The spherical mass $M$ is essential in solving EFE. When the distance is greatly increased, the Schwarzschild solution can be simplified to resemble flat Minkowski spacetime.

\item{\bf FLRW Metric}

Also, the FLRW metric gives an exact solution to the EFE. This metric characterizes a Universe that is both homogeneous and isotropic, undergoing evolution through expansion or contraction \cite{Lemaitre:1933ngn}. The basic principles of homogeneity and isotropy establish the foundational structure upon which the Standard Model of contemporary cosmology is constructed,
\begin{equation}
ds^{2}=-dt^{2}+a^{2}(t)\left(\frac{dr^{2}}{1-kr^{2}}+r^{2}d\Omega^2\right), \label{eq:FLWR1}
\end{equation}
where the cosmic time $t$ represents the proper time that comoving observers undergo while staying at rest in the comoving coordinates; $dr = d\theta = d\phi = 0$. $k$ represent the sectional curvature, where $k=0$ for flat, $k=+1$ for closed, and $k=-1$ for open curvature. The expansion or contraction of the Universe can be characterized by the scale factor $a(t)$. The scale factor dictates how the scales of the cosmic structure evolve in comparison to its initial dimensions. The substance that fills the cosmic background determine the scale factor. In the present analysis, matter-dominated background is taken into account, that is  $a(t) \propto t^{2/3}$.

\item{\bf EGS Metric}
%\label{sec:EGS1}

It is evident that the combination of the Schwarzschild vacuum metric within an evolving Universe seems to present a model that accurately represents our Universe. Einstein and Straus suggested the Swiss-cheese cosmological model during the 1940s which is sought to provide a conceptual framework for understanding the Universe as a collection of cosmic matter clusters that are nonhomogeneously distributed and interspersed with voids \cite{RevModPhys.17.120}. Gilbert proposed the successful formulation of the Schwarzschild vacuum metric within the FLRW metric \cite{Gilbert1956,Stewart:1990ngn}. In acknowledgment of their significant contributions, we wish to propose that the homogeneous solution be designated as the Einstein--Gilbert--Straus (EGS) metric.

The EGS metric is a framework to approach the nonhomogeneity of our Universe 
\bea
ds^{2}=\left(1-\frac{2M}{r}\right)dt^{2}-\frac{a(t)}{\left(1-\frac{2M}{r}\right)}dr^{2}-a(t)r^{2} d\Omega^2. \label{eq:Glbrt1}
\eea
$M$ represents a mass which is associated with a spherical distribution, for which EFE are solved. As a consequence, the spherically symmetric vacuum Schwarzschild solution is fully obtained, Eq. (\ref{eq:Glbrt1}) with $a(t)=1$. This solution describes the gravitational field that exists in the proximity of a spherical mass distribution with a total cosmic substance $M$.

As the EGS metric, Eq. \eqref{eq:Glbrt1}, indicates a double dependence on both $r$ and $t$ during the evolution of our Universe, it follows that $M$ should have the same dependence. This requires incorporating an overall expansion as the on achieved by finite cosmological constant $\Lambda$. This could be modeled by a spherical form of the Einstein-Straus vacuole \cite{PhysRev.136.B571,osti_4325043},
\bea
M(r,t) &=& \frac{4}{3} \pi f(t) r^3, \label{eq:Mrt1}
\eea 
where the arbitrary function $f(t)$ dictates the temporary evolution of the cosmic substance. The radius $r$ is accordingly adjusted along the cosmic time $t$,
\bea
f(t) &=& \frac{1}{\mu^2} \tanh\left(\frac{t}{\mu}\right). \label{eq:Mrt2}
\eea 
where the variable $\mu$ is chosen without any specific constraint rather than ensuring the appropriate mass dimension in Eq. \eqref{eq:Mrt1}. Then, the static metric described by Eq. (\ref{eq:Glbrt1}) becomes dynamic
\bea
ds^{2}=\left(1-\frac{2M(r,t)}{r}-\frac{1}{3} \Lambda r^2\right)dt^{2} - \frac{a(t)}{\left(1-\frac{2M(r,t)}{r}-\frac{1}{3} \Lambda r^2\right)} dr^{2} - a(t)r^{2} d\Omega^2, \label{eq:Glbrt2}
\eea 

\end{enumerate}

The procedure to assess the spacetime curvature and thereby to find out whether singularity or nonsingularity exists is the timelike evolution of Raychaudhuri equations, Section \ref{sec:Raychau2}. Details about the proposed geometric quantization and the related geodesic equations and affine connections shall be summarized in Appendix \ref{sec:GGA}.

\subsection{Evolution of Raychaudhuri Equations with Quantized Metric Tensor $\tilde{g}_{\alpha \beta}$}
\label{sec:Raychau2}

A set of generalized kinematic quantities, including expansion or volume scalar $\tilde{\Theta}^2$, shearing $\tilde{\sigma}^2$, rotation $\tilde{\omega}^2$, and Ricci identity $\tilde{RI}=\tilde{R}_{\mu \nu} \tilde{u}^{\mu} \tilde{u}^{\nu}$, constructs the Raychaudhuri equations. It is assumed that this is valid for both conventional and quantized metric tensor. With the quantized fundamental metric tensor $\tilde{g}_{\mu \nu}$, which can be expressed as a conformal transformation of its conventional version $g_{\mu \nu}$, Appendix \ref{sec:tildegmunuApp}, the Raychaudhuri equations read
\bea
\frac{d\tilde{\Theta}}{d\tau} &=& - \frac{1}{3} \tilde{\Theta}^2 -  \tilde{\sigma}^2 + \tilde{\omega}^2 - \tilde{RI}. \label{eq:dThetaRaych2}
\eea
The present study evaluates all these kinematic quantities in dependence on both $r$ and $t$.

\begin{enumerate} 
\item{\bf Schwarzschild Metric}

The radial geodesic assumes $\tilde{u}^{\theta}=\tilde{u}^{\phi}=0$. Then, $\tilde{u}^{t}$ and $\tilde{u}^{r}$ can be deduced as 
\bea
\tilde{u}^{t} &=& \frac{1}{F(r)} \cdot \exp\left(\frac{- 2 r K^{2}_{,\gamma}}{C(x,p)}\right) \label{eq:Schwrut2} \\
\tilde{u}^{r} &=& \pm \left[-\frac{F(r)}{C(x,p)}+\exp\left(\frac{4 K^{2}_{,\gamma}r}{C(x,p)}\right)\right]^{1/2}, \label{eq:Schwrur2}
\eea
where $C(x,p)$ represent the conformal coefficient, Eq. \eqref{eq:Ctildegalphabeta1}. In the derivation of the second velocity, the normalization condition $\tilde{g}_{\alpha\beta} \tilde{u}^{\alpha}\tilde{u}^{\beta}=-1$ is utilized. $K^2_{, \gamma}$ is the derivative of the Klein metric, Eq. \eqref{eq:Klein1}
\bea
K^{2}_{, \gamma} = \frac{\partial K^2(x_0^{\alpha}, p_0^{\beta})}{\partial x_0^{\alpha}} &=& 2 \frac{\left\langle x_0^{\alpha} \cdot p_0^{\beta}\right\rangle}{\left[\left(x_0^{\alpha}\right)^2 -1\right]^2} \left\{\left[1-\left(x_0^{\alpha}\right)^2\right] p_0^{\beta} - x_0^{\alpha} \left\langle x_0^{\alpha} \cdot p_0^{\beta}\right\rangle \right\}.
\eea
$x_0^{\alpha}$ and $p_0^{\beta}$ are auxiliary four-vectors of coordinates and momenta, respectively.

The expansion or volume scalar is determined from $\tilde{\Theta} = \tilde{u}^{\alpha}_{;\alpha}=\tilde{u}^{\alpha}_{\alpha}+\tilde{u}^{\sigma} \tilde{\Gamma}^{\alpha}_{\sigma \alpha}$, from which we find that $\tilde{\Theta} = \frac{1}{r^{2}}({r^{2}\tilde{u}^{r}})_{, r}$ and
\bea
\tilde{\Theta} &=& \pm \frac{1}{C(x,p) r^2\sqrt{\exp\left(\frac{4 r K^{2}_{,\gamma}}{C(x,p)}\right)-\frac{F(r)}{C(x,p)}}} \left\{3 M + 2 r \left[
\exp\left(\frac{4 r K^{2}_{,\gamma}}{C(x,p)}\right)\left[C(x,p)+r K^{2}_{,\gamma}\right] - 1
\right]\right\}. 
\label{eq:SchwarzTheta} \hspace*{7mm}
\eea
The evolution of $\tilde{\Theta}$ can be straightforwardly derived  
\bea
\frac{d\tilde{\Theta}}{d\tau} &=& \frac{1}{r^3 C(x,p)^2\left[2 M + r \left(C(x,p) \exp\left(\frac{4 r K^{2}_{,\gamma}}{C(x,p)}\right) -1 \right)\right]} \nn \\ 
&\times & \left\{
\frac{4 r^3 \left(K^{2}_{,\gamma}\right)^2}{C(x,p)} \exp\left(\frac{4 r K^{2}_{,\gamma}}{C(x,p)}\right)\left[4M + r \left(
C(x,p) \exp\left(\frac{4 r K^{2}_{,\gamma}}{C(x,p)}\right) -2 \right)\right] 
\right. \nn \\
& &\left. + 4 r^2 K^{2}_{,\gamma} \exp\left(\frac{4 r K^{2}_{,\gamma}}{C(x,p)}\right) \left[3 M + \left(C(x,p) \exp\left(\frac{4 r K^{2}_{,\gamma}}{C(x,p)}\right) -1 \right)\right] \right. \nn \\
& &\left. - C(x,p) \left[
\frac{9 M^2}{C(x,p)} + 8 M r \left[\exp\left(\frac{4 r K^{2}_{,\gamma}}{C(x,p)}\right)-1\right] + 2 r^2 \left[\exp\left(\frac{4 r K^{2}_{,\gamma}}{C(x,p)}\right) - 1 \right]^2 
\right]
\right\}. \label{eq:SCHWRZdTht1} \hspace*{9mm}
\eea 
We notice that the expansion and its evolution are solely depending on $r$.

\item{\bf FLRW Metric}

The FLRW metric satisfies the cosmological principles of anisotropy and homogeneity so that $\tilde{u}^r=\tilde{u}^{\theta}=\tilde{u}^{\phi}=0$ and $d \tilde{u}^t/d\tau = 0$. Then, $ \tilde{u}^t=\pm v$, where $v$ can be identified with the speed of light in vacuum. Then, the expansion is given as
\bea
\tilde{\Theta} = 3 v\left[\frac{\dot{a}(t)}{a(t)}+\frac{K^{2}_{,\gamma}}{C(x,p)}\right].  \label{eq:flrwTheta1}
\eea
The stress tensor and rotation components diminish. The Ricci identity reads
\bea
\tilde{RI} &=& v^2 \left[
\frac{1}{r - 2 r^3} + \cot(\theta) + \frac{5}{2}\frac{K^2_{, \gamma}}{C(x,p)} + 3 \frac{\dot{a}(t)^2-2 a(t) \ddot{a}(t)}{a(t)^2}\right],
\eea
The evolution equation for this type of expansion can be calculated from the Raychaudhuri equations \cite{Eric:2004ngn} 
\bea
\frac{d\tilde{\Theta}}{d\tau}&=& - v^2\left\{
\left[\frac{\dot{a}(t)}{a(t)} + \frac{K^{2}_{,\gamma}}{C(x,p)}\right]^2 + \left[
\frac{1}{r - 2 r^3} + \cot(\theta) + \frac{5}{2}\frac{K^2_{, \gamma}}{C(x,p)} + 3 \frac{\dot{a}(t)^2-2 a(t) \ddot{a}(t)}{a(t)^2}\right]\right\}.
 \label{eq:FLRWdTht1} \hspace*{9mm}
\eea
where $\dot{a}(t)=d a(t)/dt$ and $\ddot{a}(t)=d \dot{a}(t)/dt$. 
We find that the expansion and its evolution are depending on both $r$ and $t$.

\item{\bf EGS Metric}

From the quantized geodesic equations of EGS nonhomogeneous metric, we find that
\bea
\frac{d \tilde{u}^t(r,t)}{d \tau} &=& -\left[\frac{ \frac{3\de M(r,t)}{\de t}}{-3r +\Lambda r^3 + 6 M(r,t)} +\frac{F^2_{, \gamma}}{2 C(x,p)} + \frac{1}{2} K^2_{, \gamma}\right] \left(\tilde{u}^t(r,t)\right)^2,
\eea
which can be solved as
\bea
\tilde{u}^t(r,t) &=& \left[-3 + 8\pi r^2  t \tanh(t) + \Lambda r^2\right]^{-1/2} \exp\left(-\frac{[1+C(x,p)] t K^2_{, \gamma}}{2 C(x,p)}\right). \label{eq:tldu1}
\eea
The corresponding affine connections is outlined in Eq. \eqref{eq:C}. Then, the expansion or volume scalar $\tilde{\Theta}(r,t) = \tilde{u}^{\alpha}(r,t)_{,\alpha} + \tilde{u}^{\sigma}(r,t) \tilde{\Gamma}^{\alpha}_{\sigma \alpha}$ is given as
\bea
\tilde{\Theta}(r,t) &=& \frac{\exp\left[-\frac{(1+C(x,p)) t K^2_{, \gamma}}{2 C(x,p)}\right]}{2C(x,p) g(r,t)^{3/2} a(t)} \left\{
-8 \pi C(x,p) r^2 \left[t \sech^2(t)+\tanh(t)\right] a(t)\right. \nn \\
&-&\left. [C(x,p)-3] K^2_{, \gamma} g(r,t) a(t) + 3 C(x,p) g(r,t) \dot{a}(t)  
\right\}, \label{eq:Mod-theta1}
\eea
where $g(r,t)=-3 +\Lambda r^2 + 8 \pi r^2 t \tanh(t)$.

The skew part of the velocities $u_{[\alpha ; \beta]}$ represents the rotation scalar $\omega_{\alpha \beta} = u_{[\alpha ; \beta]}$. Then, the quantized quadratic rotation $\tilde{\omega}^2=\tilde{\omega}_{\mu \nu} \tilde{\omega}^{\mu \nu}$ can be given as the finite components $\omega_{12} \omega^{12}$,
\bea
\tilde{\omega}^2(r,t) &=& -\frac{C(x,p)^4 \exp\left[-2\frac{[1+C(x,p)] t K^2_{, \gamma}}{2 C(x,p)}\right]}{36 a(t) g(r,t)^{3}} \nn \\
&\times & \left\{
\Lambda r \left(-3 +\Lambda r^2\right) + 8 \pi r \tanh(t)  \left[3(t-2)+\Lambda r^2 (1+t) + 8 \pi r^2 t \tanh(t) \right]
\right\}^2.  \label{eq:Mod-omega1} \hspace*{7mm}
\eea

The stress scalar $\tilde{\sigma}_{\alpha \beta} = \tilde{u}_{(\alpha ; \beta)} - \frac{1}{3} \tilde{\Theta} \tilde{h}_{\alpha \beta}$, where $\tilde{u}_{(\alpha ; \beta)}$ are the symmetric parts of the velocities, $\tilde{h}_{\alpha \beta}=\tilde{g}_{\alpha \beta} + \tilde{u}_{\alpha} \tilde{u}_{\beta}$ is the spacial projection of the metric in directions orthogonal to the timelike vector fields $\tilde{u}_{\alpha}$, where $\tilde{u}_{\alpha} \tilde{u}^{\beta}=-1$ is the four-velocity normalization condition. 
%$\sigma^{11} \sigma_{11}$, $\sigma^{12} \sigma_{12}$, $\sigma^{22} \sigma_{22}$,  $\sigma^{33} \sigma_{33}$, and $\sigma^{44} \sigma_{44}$,
Then, the resulting nonvanishing components of $\tilde{\sigma}_{\alpha \beta}$ determine the quantized kinematic shearing  
\bea
\tilde{\sigma}(r,t) &=& \frac{C(x,p)^2 e^{-2\frac{[1+C(x,p)] r K^2_{,\gamma}}{2 (x,p)}} \csc^4(\theta)}{36 r^4 a^2(t) g(r,t)^3} \left\{K^2_{,\gamma} h(r,t) g(r,t)  \right. \nn \\
& & \left. - r^2 a(t) \sin^2(\theta) \left[8 \pi C(x,p) r^2 [t \sech^2(t)+\tanh(t)] + [C(x,p)-3] K^2_{,\gamma} g(r,t)\right] \right\}^2 \nn \\
&+&\frac{C(x,p)^2 e^{-2\frac{[1+C(x,p)] r K^2_{,\gamma}}{2 C(x,p)}}}{36 r^4 a^2(t) g(r,t)^3} \left\{K^2_{,\gamma} h(r,t) g(r,t) - r^2 a(t) \left[8 \pi C(x,p) r^2 [t \sech^2(t)+\tanh(t)] \right.\right. \nn \nn \\
&& \left.\left. + [C(x,p)-3]  K^2_{,\gamma} g(r,t)\right] \right\}^2  \nn  \\
&+& \frac{C(x,p)^2 e^{-2\frac{(1+c) r K^2_{,\gamma}}{2 C(x,p)}}}{324 a^2(t) h(r,t)^2 g(r,t)^3} \left\{
K^2_{,\gamma} h(r,t) g(r,t) \left[3(C(x,p)-3)a(t)+h(r,t)^2\right] \right. \nn \\
&&\left. + \frac{6}{r}C(x,p) a(t) \left[4 \pi r^3 \left[t \sech^2(t)+\tanh(t)\right]h(r,t)-9 g(r,t)\frac{\de M(r,t)}{\de t} \right]
\right\}^2 \nn \\
&+& \frac{C(x,p)^2 e^{-2\frac{[1+C(x,p)] r K^2_{,\gamma}}{2 C(x,p)}}}{324 h(r,t)^2 g(r,t)^3} \left\{
-\frac{h(r,t)}{a(t)} \left[-3 + C(x,p) e^{-2\frac{[1+C(x,p)] r K^2_{,\gamma}}{2 C(x,p)}} \frac{h(r,t)}{g(r,t)}\right] \times C(x,p) \right. \nn \\
&&\left.  \left[- 8 \pi  r^2 a(t) [t \sech^2(t)+\tanh(t)] - \left(1-\frac{3}{C(x,p)}\right) a(t) K^2_{,\gamma} g(r,t) + 3 g(r,t) \dot{a}(t)\right] \right. \nn \\
&& \left. + \frac{9}{r} \left[[C(x,p)+2] r K^2_{,\gamma} h(r,t) g(r,t) + 8 \pi C(x,p) r^3 \tanh(t) \right.\right. \nn \\
&& \left.\left. \hspace*{4mm}  \left[-3+r^2[\Lambda + 8 \pi \tanh(t)] + \frac{\sech(t)}{\sinh(x)}\left[(t-2)(\Lambda r^2 -3)-8 \pi r^2 t \tanh(t)\right]\right] \right. \right. \nn \\
&& \left.\left. \hspace*{4mm} + 6 C(x,p) g(r,t) \frac{\de M(r,t)}{\de t}
\right]
\right\}^2 \nn \\
&-& \frac{C(x,p)^2}{144 r^2 a(t) g(r,t)^3} e^{-2\frac{(1+C(x,p)) r K^2_{,\gamma}}{2 C(x.p)}} \left\{r K^2_{,\gamma} h(r,t) g(r,t) \right. \nn \\
&& \left. \hspace*{4mm} + 6 C(x,p) \left[ -4 \pi r^2 \tanh(t) \left(-2 + 2 t + g(r,t) \right) + g(r,t)  \frac{\de M(r,t)}{\de r}\right] \right\}^2, \label{eq:Mod-sigma1} \hspace*{0mm}
\eea
where $h(r,t)=-3 +\Lambda r^2 + 8 \pi r^2 \tanh(t)$.

In the evolution equation of the expansion, Ricci curvature tensor, $R_{\mu \nu}$ as well as Riemann and Weyl tensors appear along with the velocity field deriving the geodesic congruence. As done in the other three kinematic quantities, $\tilde{RI}=\tilde{R}_{\mu \nu} \tilde{u}^{\mu} \tilde{u}^{\nu}$ is then determined as $\tilde{RI}=\tilde{R}_{t t} \tilde{u}^{t} \tilde{u}^{t}$,
\bea
\tilde{RI}(r,t) &=& \frac{\exp\left(-4\frac{[C(x,p)-1] t K^2_{,\gamma}}{2 C(x,p)}\right)}{\left(u^t(r,t)\right)^{-2}}\left\{
\frac{1}{r} + \cot(\theta) + \frac{5 K^2_{,\gamma}}{2 C(x,p)} + \frac{3}{4} \frac{\dot{a}(t) H(t)}{h(r,t)} + 3\frac{\frac{\de M(r,t)}{\de t}}{r h(r,t)} \right. \nn \\
&+&\left.  \frac{K^2_{,\gamma}}{6 r C(x,p) a^2(t)} \left[r \dot{a}(t) h(r,t) + 6\left[-1+a(t)\right] a(t) \frac{\de M(r,t)}{\de t}\right] - \frac{18\frac{\de M^2(r,t)}{\de t}}{r^2\left[U\right]^2} \right. \nn \\
&+&\left. \frac{36 \frac{\de M^2(r,t)}{\de t} - 6 r h(r,t) \frac{\de^2 M(r,t)}{\de t^2}}{r^2 h(r,t)^{2}} + \frac{9 a(t)}{r^2 h(r,t)^3}\left[-12 \frac{\de M^2(r,t)}{\de t} + \frac{r \frac{\de^2 M(r,t)}{\de t^2}}{h(r,t)^{-1}} \right]  \right. \nn \\
&+&\left. \left[r^2\left[\Lambda - 4 \pi \tanh(t)\right] + 3 \frac{\de M(r,t)}{\de r}\right] \left[\frac{1}{r\left(-h(r,t)\right)} + \frac{[a(t)-1] K^2_{,\gamma}}{3 C(x,p) r a(t)} + \frac{1}{r h(r,t)}  \right. \right.\nn \\ 
& &\left.\left. - \frac{H(t)}{9 r a(t)} h(r,t) + \frac{2 \frac{\de M(r,t)}{\de t}}{3 r^2 a(t)} - \frac{6\frac{\de M(r,t)}{\de t}}{r^2 h(r,t)^2}
\right] + \frac{3\frac{\de^2 M(r,t)}{\de t^2}}{r h(r,t)} \right. \nn \\
&-&\left. \frac{K^2_{,\gamma}}{9 C(x,p)} \left\{r h(r,t)\left(\frac{\Lambda + 8\pi \tanh(y)}{a(t)}+\frac{36 \pi r \sech^2(t)}{h(r,t)^2}\right)  + \frac{54 \frac{\de M(r,t)}{\de t}}{r h(r,t)} \right.\right. \nn \\
&& \left.\left. + \frac{3}{r}h(r,t)\frac{\de M(r,t)}{\de t} - \frac{h(r,t)}{r a(t)} \left[\Lambda r^2 - 4\pi r^2 \tanh(t)+3 \frac{\de M(r,t)}{\de r}\right]
\right\} \right. \nn \\
&-&\left. \frac{r^2}{4 a(t)} \left[\dot{a}^2(t) - 2 a(t)\ddot{a}(t) + \frac{6 a(t) \dot{a}(t) \frac{\de M(r,t)}{\de t}}{r h(r,t)} \right.\right. \nn \\
&& \left.\left. + \frac{4 a(t)}{9 r^2} h(r,t) \left[\Lambda r^2 - 4\pi r^2 \tanh(t)+3 \frac{\de M(r,t)}{\de r}\right] 
\right] \right. \nn \\
&-&\left. \frac{r^2 \sin^2(\theta)}{4 a(t)}\left[\dot{a}^2(t) - 2 a(t) \ddot{a}(t) + \frac{6 a(t) \dot{a}(t) \frac{\de M(r,t)}{\de t}}{r h(r,t)} \right.\right. \nn \\
&& \left.\left. + \frac{4 a(t)}{9 r^2} h(r,t) \left[\Lambda r^2 - 4\pi r^2 \tanh(t)+3 \frac{\de M(r,t)}{\de r}\right] 
\right] \right. \nn \\
&+&\left. \left[r\frac{\de^2 M(r,t)}{\de r \de t}-\frac{\de M(r,t)}{\de t}\right] \left[\frac{3}{r^2 h(r,t)} + \frac{h(r,t)}{3 r^2 a(t)}\right] \right. \nn \\
&+&\left. \frac{1}{r^2 h(r,t)^2}\left[18\frac{\de M(r,t)}{\de t}\left(-1+\Lambda r^2 + 2 \frac{\de M(r,t)}{\de r}\right) - 6r h(r,t) \frac{\de^2 M(r,t)}{\de r \de t}\right] \right. \nn \\
&+& \left. \frac{\tanh^3(t)}{6 r^5 h(r,t)^2}\left[1024 \pi^3 r^9 + \frac{384 \pi^2 r^7}{\tanh(t)}\left(-2+\Lambda r^2 - 2 \frac{\de M(r,t)}{\de r} + r \frac{\de^2 M(r,t)}{\de r^2}\right) \right.\right. \nn \\
&&\left.\left. - \frac{24 \pi r^5}{\tanh^2(t)} \left[3 r^2 \ddot{a}(t) + 2\left(\Lambda r^2-3\right)\left(1-\Lambda r^2 + 4 \frac{\de M(r,t)}{\de r} - 2 r \frac{\de^2 M(r,t)}{\de r^2}\right) \right] \right.\right. \nn \\
&&\left.\left. + r^3 \left[-9 r^2\left(\Lambda r^2-3\right)\ddot{a}(t) + 81 r \dot{a}(t) \frac{\de M(r,t)}{\de t} \right. \right. \right. \nn \\
&&\left.\left.\left. \hspace*{9mm} + 2 \left(\Lambda r^2-3\right)^2 \left(-6\frac{\de M(r,t)}{\de r}+r \left[\Lambda r + 3 \frac{\de^2 M(r,t)}{\de r^2}\right]\right)
\right]
\right]
\right\}. \label{eq:Mod-R11u11}
\eea

An analytic expression for the evolution of the quantized Raychaudhuri equations can be combined from Eqs. (\ref{eq:Mod-theta1}), (\ref{eq:Mod-omega1}), (\ref{eq:Mod-sigma1}), and (\ref{eq:Mod-R11u11}).

\end{enumerate} 

The numerical results from the three metrics shall be elaborated in the section that follows.

\section{Numerical Results and Discussion}
\label{sec:dsct}

The formulation of the Raychaudhuri equation, as expressed in Eq. (\ref{eq:dThetaRaych2}), is established through the quadratic invariants of various geometric quantities: the volume scalar (expansion), Eq. (\ref{eq:Mod-theta1}), the rotation (spinning), Eq. (\ref{eq:Mod-omega1}), the shearing (anisotropy), Eq. (\ref{eq:Mod-sigma1}), and the Ricci identity (local gravitational field), Eq. (\ref{eq:Mod-R11u11}). This equation characterizes the evolution of expansion in terms of proper time, framed by the velocity fields $\tilde{u}^t(r,t)$. In examining the implications of the local gravitational field, as encapsulated by the Ricci tensor, we have purposefully constructed the term $\tilde{R}_{\mu \nu} \tilde{u}^{\mu} \tilde{u}^{\nu}$ from foundational concepts. One might represent $\tilde{R}_{\mu \nu} \tilde{u}^{\mu} \tilde{u}^{\nu}$ as $(\rho + 3p)/2 - \Lambda$, where $\rho$ refers to the density of the cosmic substance and $p$ indicates its isotropic pressure. Our decision to avoid the integration of thermodynamic variables is motivated by the intention to reduce, or ideally eliminate, any potential model-dependence.

The proper-time derivative within the $u^t(r,t)$-frame is formulated as $\dot{\Theta}(r,t)=u^t(r,t) \nabla_t \Theta(r,t)$, and the trace of the resulting expression consists of the following components. 
\begin{enumerate}
\item The scalar quantity that quantifies the fractional rate of change in the volume of a small cluster of cosmic material over time, as observed by a central co-moving observer, represents the evolution equation governing the expansion of the timelike geodesic congruence. 
\item Vorticity that quantifies the inclination of adjacent world lines to spiral around each other, results in the rotation of the volume of cosmic matter.
\item Shearing that refers to the phenomenon that quantifies the inclination of a spherical volume of cosmic material to undergo deformation into an ellipsoidal configuration.
\item The Ricci identity that accounts for the influences of the local gravitational field.
\end{enumerate}

Before we proceed to the numerical results, it is important to highlight several considerations regarding the numerical estimation of $C(x,p)$ as indicated in Eq. \eqref{eq:Ctildegalphabeta1}. Notably, the current representation of $C(x,p)$ is a truncated expression. To facilitate the numerical analysis of the evolution of Raychaudhuri equation and to ascertain whether a big bang or bounce transpires in the curvature of spacetime, scalar values must be assigned to the conformal coefficient $C(x,p)$. This methodology can be characterized as a variant of mean-field approximation. It is essential to recognize that the qualitative numerical estimation is probably affected by an averaging mechanism. Typically, an approximation is inherently permissible, particularly in light of the significant truncation applied to the fundamental tensor, as detailed in Appendix \ref{sec:tildegmunuApp}. Conversely, while the suggested averaging process may appear to favor a particular instance within the continuous spectrum of quantum operators, it is unlikely to substantially alter the overall qualitative numerical assessment. This procedure assures the maintenance of coherence within the conformal coefficient. It underscores the prominent "measurement problem" associated with quantum mechanics. Nonetheless, this situation is an unavoidable outcome of the collapse of quantum features that occurs during the measurement process. It is essential to clarify that the calculations involved are based on assumed values for the derivative of the Klein metric $K^2_{,\gamma}$, $\Lambda$, the arbitrary parameter $\mu$, and the mass $M$, which are all assigned a value of $0.95$. With respect to $C(x,p)$, we evaluate the results at $C(x,p)=0.75$ and $C(x,p)=1.25$. The first value denotes the nonexistence of quantization, in contrast to the latter value, which is linked to geometric quantization.

\subsection{Schwarzschild Metric with $g_{\alpha \beta}$ and $\tilde{g}_{\alpha \beta}$}
\label{sec:Schwrz-dsct}
 
%%%%%%%%%%%%%%%%%%%%%%%%%%%%%%%%%%%%%%
\begin{figure}[htb!] 
\includegraphics[angle=-90,width=0.7\textwidth]{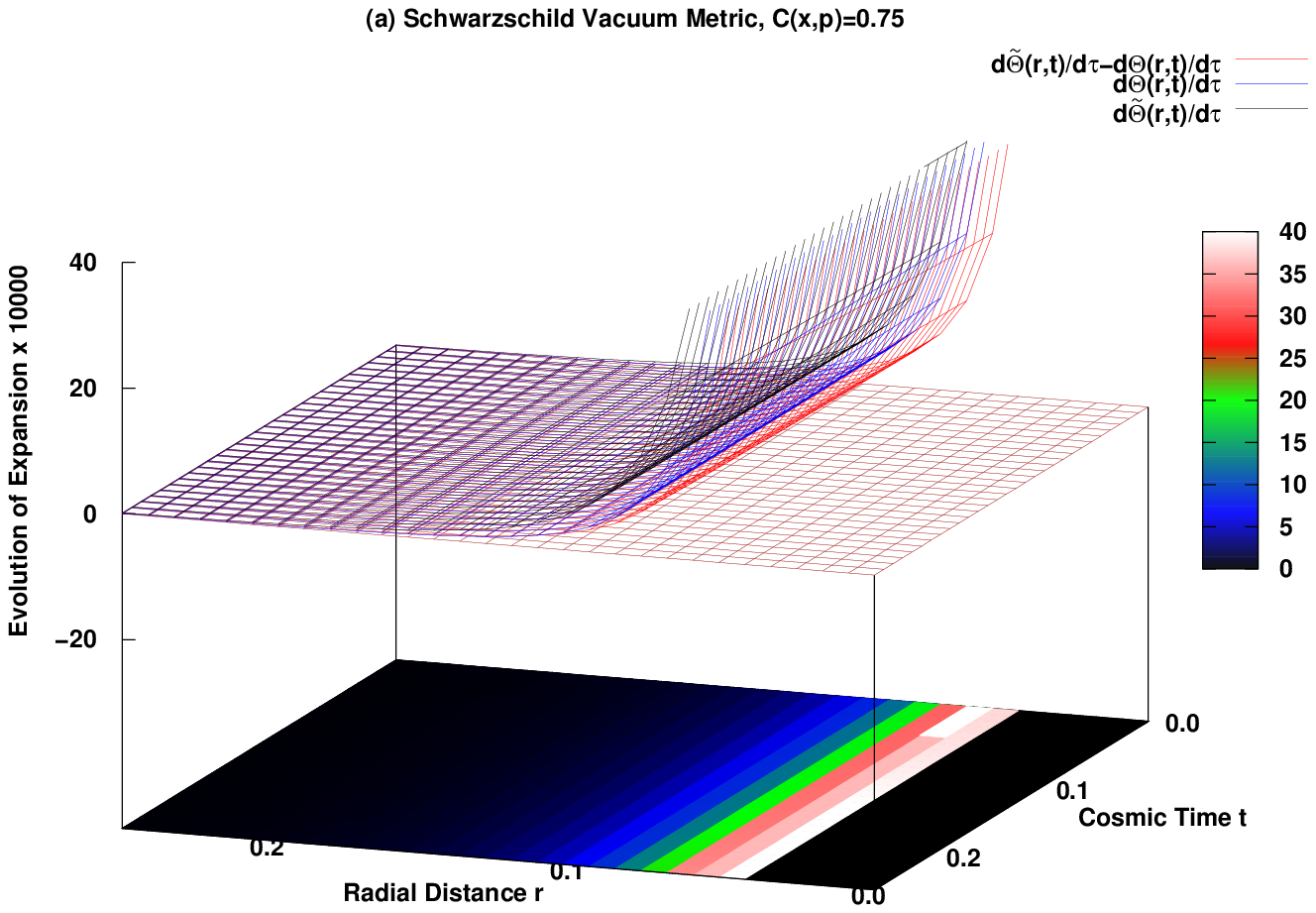}\\
\includegraphics[angle=-90,width=0.7\textwidth]{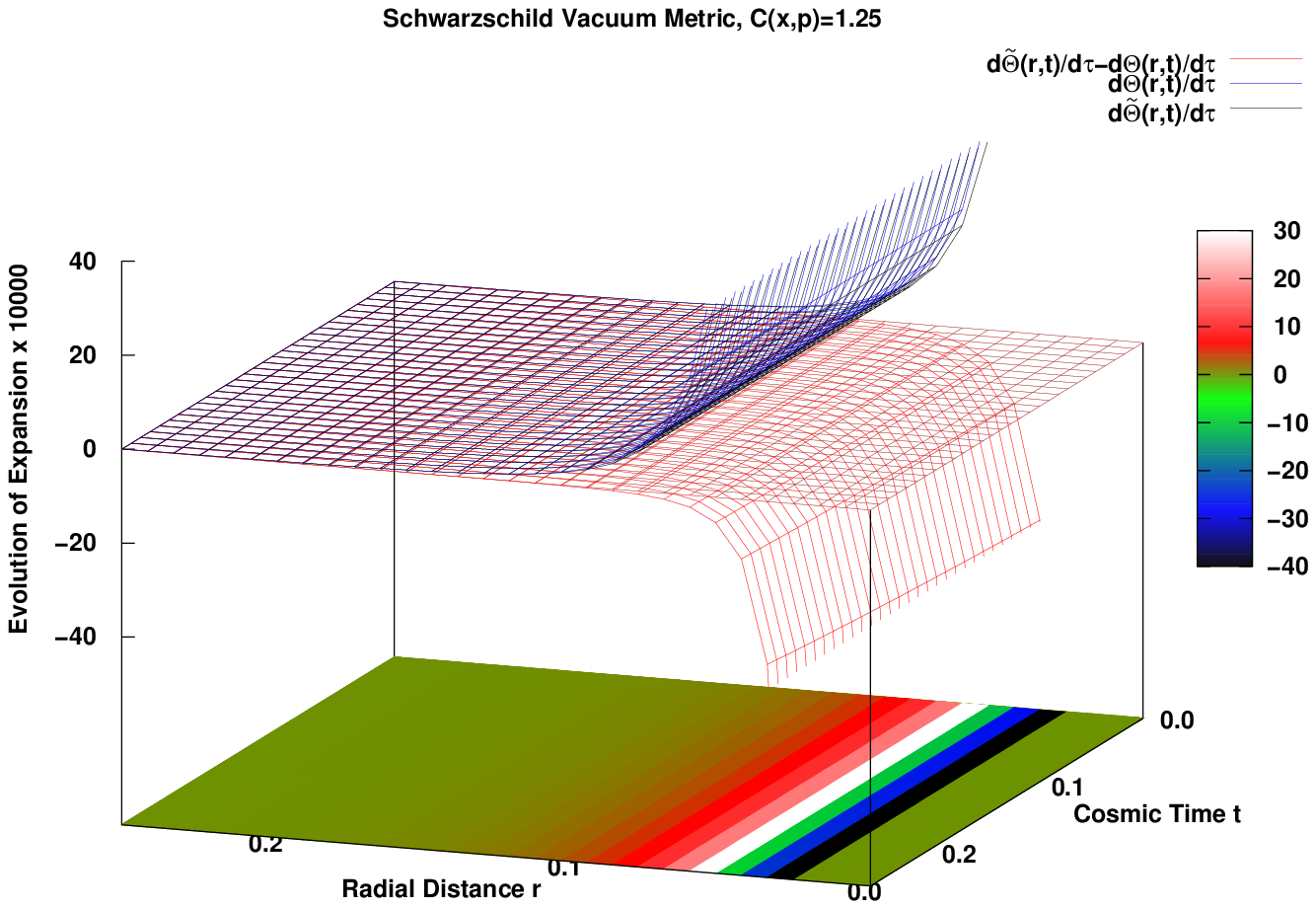}  
\caption{The evolution of Raychaudhuri equations for the Schwarzschild metric is illustrated as a function of both $r$ and $t$. A comparison is made between the classical results and those derived from geometric quantization. The outcomes obtained using the conventional metric tensor, without any quantization, represented as $d\Theta/d\tau$, are shown as a blue lattice. The quantum contributions, expressed as $d\tilde{\Theta}/d\tau - d\Theta/d\tau$, are depicted by a red lattice. The findings from both conventional and quantized metric tensors are illustrated using a black lattice. Additionally, the gray lattice represents the evolution of expansion along straight world-lines, indicating the absence of curvature. The upper panel displays results at $C(x,p)=0.75$, while the lower panel shows results at $C(x,p)=1.25$.  \label{fig:1} }
\end{figure}
%%%%%%%%%%%%%%%%%%%%%%%%%%%%%%%%%%%%%%

The analytic representations associated with the Schwarzschild homogeneous metric are provided in Eq. (\ref{eq:SCHWRZdTht1}), and the relevant numerical evolution is shown in Fig. \ref{fig:1}. A comparison is made between the results obtained for $g_{\alpha \beta}$ and those for $\tilde{g}_{\alpha \beta}$. The two values enable us to evaluate the effects of quantization both below and above unity, where the proposed geometric quantization is no longer applicable. 

The analysis of results obtained from $g_{\alpha \beta}$ and $\tilde{g}_{\alpha \beta}$ facilitates the categorization of the evolution of Raychaudhuri equations into three primary types: i) classical contributions, represented by $d\Theta/d\tau$ (blue lattice), ii) quantization contributions, expressed as $d\tilde{\Theta}/d\tau - d\Theta/d\tau$ (red lattice), and iii) a hybrid of both contributions, characterized by $d\tilde{\Theta}/d\tau$ (black lattice). There exists a notable distinction in the results when comparing $C(x,p)=0.75$ and $C(x,p)=1.25$, especially regarding the pure quantized evolution of the Raychaudhuri equations. As a reference to flatness, i.e., zero curvature, the results from straight world-lines are illustrated by the gray lattice.  

According to the focusing theorem, the negative evolution of $\Theta$ indicates a contraction occurring during the evolution of the geodesic congruence. As $\Theta>0$ indicates a diverging congruence, its divergence diminishes over time, in contrast to the converging congruence at $\Theta<0$, which tends to converge at an increased pace. These findings suggest a physical interpretation wherein the significant gravitational attraction results in the focusing of geodesics. The temporal changes in spacetime surrounding a segment of matter, as observed by a central comoving observer, can be analyzed using the evolution equation related to expansion, which describes the expansion rate of the timelike congruence.

In certain world lines, the evolution with respect to the proper time $\tau$, which corresponds to the cosmological time $t$, seems to exhibit either a negative or positive value due to the values assigned to $C(x,p)$. A negative value suggests that the expansion is likely to be succeeded by a collapse, whereas a positive value indicates that the expansion persists. This analysis highlights the dynamics in the vicinity of a matter segment, as described by the Landau--Raychaudhuri equation \cite{Raychaudhuri1957,Landau:1975pou,Diab:2016dzv}, which is a fundamental lemma for the Penrose--Hawking singularity theorems concerning classical black holes \cite{Penrose1965} and the entire Universe \cite{hawking1973large}.

We also notice that the degree of quantization plays a significant role. At $C(x,p)=0.75$, the effects of pure quantization are positive throughout, whereas at $C(x,p)=1.25$, they are entirely negative. From Fig. \ref{fig:1}, a crucial conclusion can be derived that the Schwarzschild solution of EFE is likely associated with a predominant spatial singularity, i.e., a minimal initial singularity presents.

\subsection{FLRW Metric with $g_{\alpha \beta}$ and $\tilde{g}_{\alpha \beta}$}
\label{sec:RESchFLRW}

%%%%%%%%%%%%%%%%%%%%%%%%%%%%%%%%%%%%%%
\begin{figure}[htb!] 
\includegraphics[angle=-90,width=0.7\textwidth]{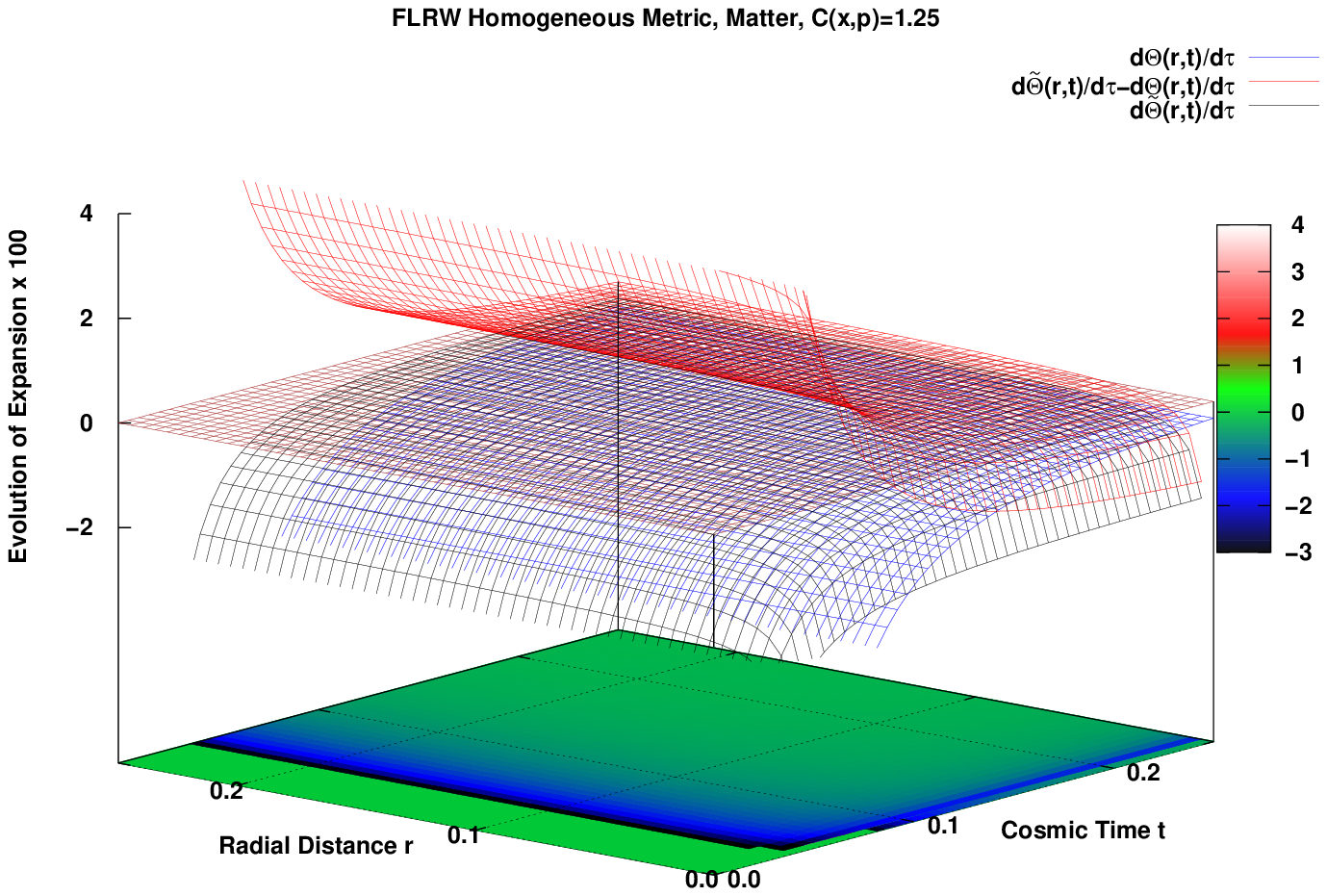} 
\includegraphics[angle=-90,width=0.7\textwidth]{dthetadtau-ClassicalQuantum-FLRW-Matter1a.eps} 
\caption{The same as in Fig. \ref{fig:1} but here of FLRW metric. The gray lattice signifies a state of flatness or zero curvature, indicating a uniform absence of evolution as described by the Raychaudhuri equations. \label{fig:2}}
\end{figure}
%%%%%%%%%%%%%%%%%%%%%%%%%%%%%%%%%%%%%%

Our numerical analysis, as shown in Figure \ref{fig:2}, details the evolution of the Raychaudhuri equations for the FLRW metric, utilizing both $g_{\alpha\beta}$ and $\tilde{g}_{\alpha\beta}$. The results indicate a clear orthogonality between the outcomes of this metric and those obtained from the Schwarzschild metric, illustrated in Figure \ref{fig:1}. While the conventional EFE yield a positive evolution of the Raychaudhuri equations, the contributions from the proposed geometric quantization result in a stark reversal of this positivity. Therefore, the implications of the proposed quantization are notably impactful. Moreover, the analysis presented in Fig. \ref{fig:1} indicates that the FLRW solution is linked to a prominent initial singularity. There is virtually no presence of spatial singularities in conjunction with this. An additional noteworthy finding is that the extent of quantization, whether at $C(x,p)=0.75$ or $C(x,p)=1.25$, seems to exert little effect on the results. At both values, the contributions from pure quantization are consistently positive and remain unchanged throughout.  

The analysis of the Schwarzschild metric, which highlights geodesic congruence in spatial dimensions, reveals that the FLRW metric is more oriented towards the temporal dimension. This leads to a negative evolution of the Raychaudhuri equations with respect to comoving time in the FLRW solution. However, the process of quantization flips this outcome to a positive value. The gray lattice is utilized as a reference for straight world-lines. A similar contrasting situation is found in the Schwarzschild solution. In both instances, quantization aims to transform divergence into convergence and vice versa.
 
The analysis of both the Schwarzschild and FLRW solutions leads to the conclusion that the proposed quantization demands a thorough reconsideration of the lessons learned from the classical approximation of general relativity. The application of geometric quantization will clarify the complementary aspects of these solutions.

\subsection{EGS Metric with $g_{\alpha \beta}$ and $\tilde{g}_{\alpha \beta}$ }
\label{sec:NrRsltTildaTheta}

%%%%%%%%%%%%%%%%%%%%%%%%%%%%%%%%%%%%%%
\begin{figure}[htb!] 
\includegraphics[angle=-90,width=0.7\textwidth]{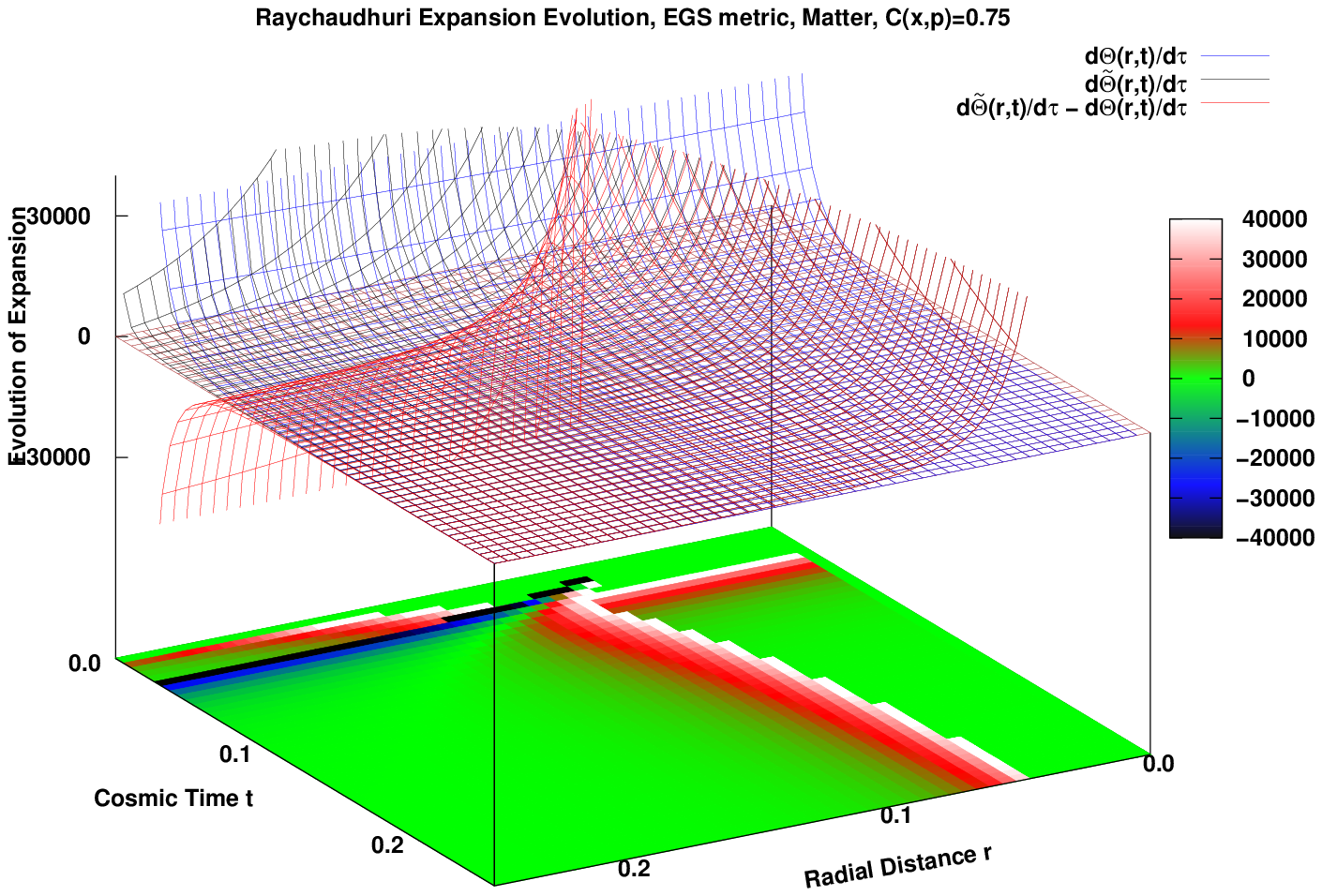}  
\includegraphics[angle=-90,width=0.7\textwidth]{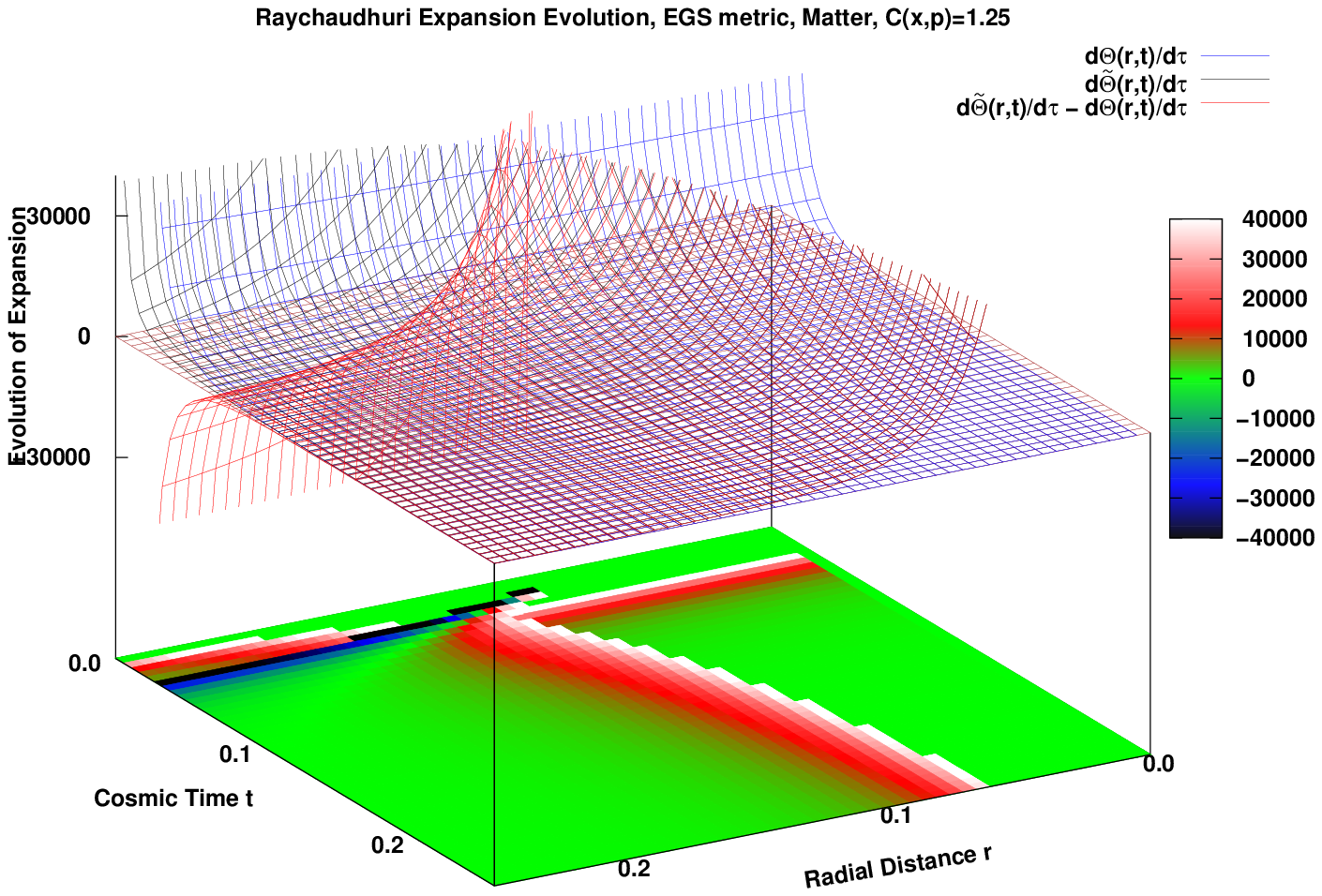} 
\caption{The same as in Fig. (\ref{fig:1}) and Fig. (\ref{fig:2}) but here for the EGS metric. \label{fig:3}}
\end{figure}
%%%%%%%%%%%%%%%%%%%%%%%%%%%%%%%%%%%%%%

The numerical findings associated with the EGS metric are depicted in Fig. \ref{fig:3}. Further insights into this metric can be found in Section \ref{sec:frmlsm0}. It is noteworthy that the evolution of the Raychaudhuri equations does not appear to be significantly influenced by cosmic time. This limitation, which suggests a near-exclusive dependence on radial distance, may stem from the dominance of FLRW-type contributions in the EGS metric. In contrast, the contributions from the Schwarzschild-type are likely to be considerably underestimated. This discrepancy may elucidate why the EGS metric, despite its accurate modeling of our Universe, has not become predominant among cosmological researchers.

Alternatively, the introduced geometric quantization approach evidently restores the simultaneous relationship of the EGS metric with both radial distance and cosmic time. This numerical analysis, based on the geometric quantization of general relativity, elucidates the pivotal role of the EGS metric in the modeling of the Universe's expansion and evolution. The hybrid results indicate that the evolution of the Raychaudhuri equations experiences a substantial increase with the reduction of both $r$ and $t$.

This critical difference becomes apparent when analyzing the numerical data derived from the pure quantum contributions (indicated by the red lattice). A reduction in the variable $t$ results in a clear upward trend in the results, while a decrease in $r$ corresponds to a downward trend in the negative direction. The complex interplay between $r$ and $t$ distinguishes this model from the classical approximation of general relativity. Furthermore, it enhances the utility of the EGS metric, thereby validating its initial conception as an effective tool for realistically modeling our Universe.

Due to the significant structure of the Raychaudhuri equations, the various dynamical quantities contributing to the numerical results of their evolution concerning comoving time will be analyzed in Appendix \ref{App:EGA-Dtls}. These factors encompass the volume scalar (expansion), rotation (spinning), shearing (anisotropy), and the Ricci identity (local gravitational field), with their analytical formulations represented in Eq. (\ref{eq:Mod-theta1}), Eq. (\ref{eq:Mod-omega1}), Eq. (\ref{eq:Mod-sigma1}), and Eq. (\ref{eq:Mod-R11u11}).

In this regard, it is noteworthy that a conjectured approach to prevent, at least in the initial stages, the occurrence of singularity has been proposed, which combines a nonvanishing cosmological constant, nonvanishing spin (rotation), and nonvanishing shearing (anisotropy). This approach appears to be i) specifically formulated for the simplest solutions of the EFE, such as the Schwarzschild and FLRW metrics, and ii) not suitable for the EGS solution. To the best knowledge of authors, there is no literature for the evolution of Raychaudhuri equations in with EGS metric. This investigation is the first of its kind in the literature. The nonvanishing cosmological constant, along with rotation and shearing, introduces aspects of expansion, rotation, and anisotropy, respectively. However, the possibility of recollapse appears to be unfavorable. The evolution equation concerning expansion in the EGS solution of the EFE is primarily determined by the volume scalar (expansion) and shearing (anisotropy). While their contributions are relatively minor, both vorticity (rotation) and the Ricci identity (local gravitational field) also support the concept of nonsingularity.

In conclusion, it is important to emphasize that the evolution of the Raychaudhuri equations reveals that the fundamental geometric properties of spacetime must be distinctly defined. The numerical results presented in Fig. \ref{fig:3} do not contradict the principles underlying the Raychaudhuri evolution equations, which serve as a nonsingular model of the Universe. This model was initially established for the most simplest vacuum solution, and the most plausible nonhomogeneous solution appears to be consistent with this framework.

\section{Conclusions}
\label{sec:cncl}

The present study employs analytical and numerical approaches to assess spacetime curvatures in relation to radial distance and cosmic time. By assuming the cosmic background can be characterized as a vacuum or as homogeneous or nonhomogeneous, we compare the evolution of the Raychaudhuri equations with both conventional and quantum metric tensors. To achieve this objective, we systematically compare the Schwarzschild, FLRW, and EGS solutions of EFE through both analytical and numerical analyses. The nonhomogeneous background effectively illustrates a realistic model of our Universe. The distribution of substance is nonhomogeneously spread across the voids and holes present in the structure. It is hypothesized that the volume of substance contained within these lumps corresponds to the amount removed to create the voids. The Swiss-cheese cosmological model, first introduced by Einstein and Straus and subsequently enhanced by Gilbert, serves as a realistic framework for modeling our Universe. This model, referred to as the EGS metric, represents an extension of the conventional cosmological model, particularly addressing the implications of nonhomogeneity in the context of an expanding Universe.

Utilizing both the conventional and quantized fundamental metric tensors, this analysis explores the evolution of the Raychaudhuri equations with respect to comoving time within the contexts of Schwarzschild, FLRW, and EGS metrics, thereby characterizing the spacetime curvatures. The integration of a quantum geometric approach and a kinematic theory of free-falling quantum particles on an extended tangent bundle manifold results in quantum-mechanically induced alterations to the associated metric. Consequently, a quantized fundamental metric tensor emerges, which can be described as a conformal transformation of the conventional metric. The mean-field approximation enables effective averaging of the conformal coefficient, thus facilitating the calculations related to the evolution of the Raychaudhuri equations. 

The numerical analysis concerning the evolution of Raychaudhuri equations sheds light on the three metrics that model vacuum, homogeneous, and nonhomogeneous cosmic backgrounds. By comparing the numerical results from the conventional metric tensor with those derived from the quantum metric tensor, one can discern the implications of the proposed geometric quantization approach.

The analysis of the Schwarzschild metric reveals that the evolution of the Raychaudhuri equations is predominantly positive, especially in the regime of low $r$. It is important to note that this evolution shows no correlation with cosmic time $t$. Upon varying the conformal coefficient, the contributions arising from pure quantum effects are observed to be negative, particularly at lower $r$. In contrast, the hybrid results continue to display a positive nature.

The volution of Raychaudhuri equations with respect to comoving time, when analyzed within the framework of the FLRW metric, demonstrate a negative evolution with respect to both conventional and hybrid metric tensors. In contrast, the contributions from pure quantum effects are uniformly positive throughout. Furthermore, variations in the conformal coefficient do not alter the invariance of these results.

The conventional metric tensor associated with the EGS metric appears to restrict the evolution of the Raychaudhuri equations to changes in time, leading to a largely positive result for small time intervals. Interestingly, there is no observable dependence on the radial coordinate, despite the initial assumption that the EGS metric would integrate features of both the Schwarzschild and FLRW metrics. With the implementation of the proposed quantization, however, the EGS metric retains its expected variation with respect to both the radial coordinate and time. Consequently, we conclude that the proposed geometric quantization offers a crucial extension to the EGS solution of the Einstein Field Equations, confirming that the EGS metric is indeed influenced by both the radial coordinate and time.
 
From the analysis presented in this study, we can derive several conclusions: i) the Raychaudhuri evolution equation is capable of modeling both singular and nonsingular universes, ii) it appears that neither the Schwarzschild nor the FLRW metrics adequately or collectively contribute to the EGS metric, indicating that the EGS metric cannot simply be regarded as a combination of these metrics, iii) the nature of the EGS metric seems to diverge from that of both the Schwarzschild and FLRW metrics, iv) the singularity-free modeling of our nonhomogeneous universe, as articulated by the EGS metric, appears to be independent of specific models, v) finally, further exploration of the EGS solution of the Einstein field equations (EFE) is warranted in subsequent studies. Further implications might be investigated  elsewhere \cite{Chokyi:2024nis}.

\newpage

\appendix

\section{Quantized Fundamental Metric Tensor $\tilde{g}_{\alpha\beta}$}
\label{sec:tildegmunuApp}

The relativistic generalized uncertainty principle (RGUP) exhibits a key feature of quantum mechanics by introducing uncertainty into four-dimensional spacetime \cite{NasserTawfik:2024afw,Tawfik:2023onh,Tawfik:2023ugm,Tawfik:2023hdi,Tawfik:2024gtg,Farouk:2023hmi,Tawfik:2023orl}. In RGUP, the momentum $p_0^{\nu}$ undergoes a deformation to $\phi(p_0) p_0^{\nu}$, where $\phi(p_0) = 1 + 2 \beta p_{0}^{\rho}p_{0 \rho}$, with $\beta$ representing the RGUP parameter and $p_0$ denoting the auxiliary four-vector of momentum for a free-falling quantum particle with positive mass $m$. Additionally, in order to incorporate the kinematics of the free-falling quantum particle based on its positive homogeneity, the Finsler (Hamilton) structure can also be extended. The extended Finsler (Hamilton) structure is discretized due to RGUP and  characterized by coordinates $x_0^{\alpha}$ and direction of $\phi(p_0) p_0^{\beta}$. Through a quantum geometrical approach, the four-dimensional fundamental tensor on a Riemannian manifold can be derived from Finslerian metric, which includes quantum-mechanically induced modifications \cite{Tawfik:2023hdi,Tawfik:2023ugm,Tawfik:2023rrm,Tawfik:2023kxq} 
\bea
\tilde{g}_{\mu \nu}  &=& \left(\phi(p_0)^2+2\frac{\kappa}{(p_0^0)^2} K^2\right) 
 \left[1 + \frac{\dot{p}_0^{\mu} \dot{p}_0^{\nu}}{{\mathscr F}^2} \left(1+2\beta p_0 ^{\rho} p_{0 \rho}\right) \right] g_{\mu\nu} \nn \\
&+& \left[\frac{d x_0^{\mu}}{d \zeta^{\mu}} \frac{d x_0^{\nu}}{d \zeta^{\nu}} +  \left(1+2\beta p_0 ^{\rho} p_{0 \rho}\right) \frac{d p_0^{\mu}}{d \zeta^{\mu}} \frac{d p_0^{\nu}}{d \zeta^{\nu}} \right] d_{\mu\nu}(x,p),  \label{eq:gmunuQ2}
\eea
where the quantity ${\mathscr F}$ represents the maximal proper force \cite{Tawfik:2023kxq,Tawfik:2023orl,Tawfik:2023rrm,Farouk:2023hmi,Tawfik:2021ekh}, which corresponds to the maximum proper acceleration discovered by Caianiello \cite{Caianiello:1981jq,caianiello1984maximal,brandt1989maximal} and experienced by the free-falling particle in a curved spacetime.  The motion of a free-falling quantum particle is governed by ${\mathscr F}$, which is impacted by the additional curvatures introduced through the proposed quantization process. This connection is established through the association with the maximal proper acceleration ${\mathscr A}=2\pi(c^7/G\hbar)^{1/2}$. The appearance of ${\mathscr F}$ and ${\mathscr A}$ as novel physical constants is a consequence of the quantum geometric approach to curved spacetime \cite{Rosen:1962mpv,Caianiello:1989wm,Caianiello:1989pu}.

$\dot{p}_0^{\mu} \dot{p}_0^{\nu}$ and $\frac{d p_0^{\mu}}{d \zeta^{\mu}} \frac{d p_0^{\nu}}{d \zeta^{\nu}}$ represent the actual force responsible for the actual acceleration experienced by the free-falling quantum particle. It is clear that $\dot{p}_0^{\mu} \dot{p}_0^{\nu}/{\mathscr F}^2 \leq 1$. The same is also valid for $\frac{d p_0^{\mu}}{d \zeta^{\mu}} \frac{d p_0^{\nu}}{d \zeta^{\nu}}/{\mathscr F}^2$.
Analogous to the indexes $\alpha$ and $\beta$, the indexes $\mu, \nu \in \{0,1,2,3\}$ are treated as free indexes. The variable $\zeta^{\mu}$ serves as a parametrization that establishes a link between the coordinates in the Finslerian tangent bundle and the Riemann coordinates. It is of utmost importance to highlight that in order to ensure the equivalence of the Finslerian and Riemannian measures of the line element, the expression for $d_{\mu\nu}(x,p)$ in the second line of Eq. \eqref{eq:gmunuQ2}
\bea
d_{\mu\nu}(x,p) &=&  \frac{6 \kappa \phi(p_0)}{(p_0^0)^2}\left\{K^2 \ell_{\mu} \ell^{\sigma} g_{\sigma \mu} - K^3 \ell^{\sigma} \left[\delta_{0 \nu} g_{\sigma\mu} + \delta_{0 \mu} g_{\sigma\nu}\right] 
+ \frac{2+\phi(p_0)}{8 \phi(p_0)} K^4 \delta_{0 \mu} \delta_0^{\sigma} g_{\sigma\nu}\right\}, 
\eea 
needs to be redefined in terms of $g_{\mu\nu}(x,p)$. This redefinition is necessary to maintain consistency with the first line of Eq. \eqref{eq:gmunuQ2}, where the measures are already identical. Until the resolution of this mathematical challenge, it is necessary to make a rough truncation that the second line disappears. However, it is important to note that the function $\phi(p_0)$ is independent on $x_0$ indicating that $\phi(p_0)$ possesses a finite value. This finite value of $\phi(p_0)$ imposes certain limitations on the proposed truncation. In other words, finite function $\phi(p_0)$ opposites vanishing $d_{\mu\nu}(x,p)$ or $d_{\mu\nu}(x,p)=0$ requires vanishing $\phi(p_0)$.

In order to thoroughly examine the potential consequences of the proposed quantization approach in proving or disproving the existence of space and initial singularity, and in the absence of any viable alternative to the non-truncated $\tilde{g}_{\mu \nu}$, Eq. \eqref{eq:gmunuQ2}, it becomes inevitable to truncate it to \cite{Tawfik:2023hdi,Tawfik:2023ugm}: 
\bea
\tilde{g}_{\mu \nu}  &=& \left(\phi(p_0)^2+2\frac{\kappa}{(p_0^0)^2} K^2\right) 
 \left[1 + \frac{\dot{p}_0^{\mu} \dot{p}_0^{\nu}}{{\mathscr F}^2} \left(1+2\beta p_0 ^{\rho} p_{0 \rho}\right) \right] g_{\mu\nu} = C(x,p) g_{\mu\nu}. \label{eq:gmunuQ3}
\eea 
Incorporating the proposed quantum-mechanical revisions, the conformal coefficient $C(x,p)$ of $g_{\mu\nu}$ exhibits a clear dependence on both $x$ and $p$, i.e., phase-space dimension. 

The function $\phi(p_0)=1+\beta p_0^{\rho} p_{0\rho}$ represents another significant discovery.
\begin{itemize}
\item By incorporating the function $\phi(p_0)$, the Heisenberg uncertainty principle (HUP) is extended to the Relativistic Generalized Uncertainty Principle (RGUP). This expansion enables the examination of the effects of relativistic energy and finite gravitational fields on quantum mechanics \cite{Tawfik:2021ekh}.
\item Through the quantization of the Finsler metric tensor, $\phi(p_0)$ introduces a generalization of the fundamental tensor, thereby expanding the framework of general relativity (GR) \cite{Tawfik:2023rrm,Tawfik:2023kxq}. 
\item By maintaining the unique curvature properties of the Randers metric, $\phi(p_0)$ effectively combines gravity and electromagnetism into a unified framework \cite{Cheng2012},
\bea
\phi(p_0) &=& 1 + \beta p_0^{\rho} p_{0 \rho} = 1 + \frac{\kappa}{(p_0^0)^2} K^2.
\eea
\end{itemize} 

In the context of Klein metric $K$, the coordinates and momenta of a free-falling particle with a finite positive mass $m$ are represented by auxiliary four-vectors, namely $x_0^{\alpha}$ and $p_0^{\beta}$, respectively \cite{Klein1910},
\bea
K^2(x_0^{\alpha}, p_0^{\beta}) &=& \frac{\left\| p_0^{\beta}\right\|^2 - \left\|x_0^{\alpha}\right\|^2 \left\| p_0^{\beta}\right\|^2 + \left\langle x_0^{\alpha} \cdot p_0^{\beta}\right\rangle^2}{1-\| x_0^{\alpha}\|^2}. \label{eq:Klein1}
\eea
The standard Euclidean norm and inner product in $\R^n$ are denoted by $\|\cdot\|$ and $\langle \cdot\rangle$, respectively. In order to maintain $0$-homogeneity of $\phi(p_0)$, the RGUP method can be directly utilized for the $1$-homogeneous function $F(x_0^{\alpha}, p_0^{\beta})$ with respect to $p_0^{\beta}$.
\bea
K^2(x_0^{\alpha}, \phi(p_0) p_0^{\beta}) &=& \frac{\left\| \phi(p_0) p_0^{\beta}\right\|^2 - \left\| x_0^{\alpha}\right\|^2 \left\| \phi(p_0) p_0^{\beta}\right\|^2 + \left\langle x_0^{\alpha} \cdot \phi(p_0) p_0^{\beta}\right\rangle^2}{1-\left\| x_0^{\alpha}\right\|^2} = 
\phi(p_0)^2 K^2(x_0^{\alpha}, p_0^{\beta}). \label{eq:Klein2}
\eea
The quantized Finsler metric is obtained by considering the Hessian matrix of the squared $K(x_0^{\alpha}, \phi(p_0) p_0^{\beta})$
\bea
\tilde{g}_{\alpha \beta}(x,p)&=&\left(\phi(p_0)^2+2\frac{\kappa}{(p_0^0)^2} K^2\right) g_{\alpha \beta}(x,p) \nn \\
&+& 2 \frac{\kappa}{(p_0^0)^2} \left[4 \phi(p_0) K^2 \ell_{\alpha} \ell^{\sigma} g_{\sigma \alpha}(x) - 4 \phi(p_0) K^3 \ell^{\sigma} \left[\delta_{0 \beta} g_{\sigma \alpha}(x,p) + \delta_{0 \alpha} g_{\sigma \beta}(x,p)\right] \right. \nn \\
&&\left. \hspace*{9mm} + K^4(2+\phi(p_0)) \delta_{0 \alpha} \delta_0^{\sigma} g_{\sigma \beta}(x,p) \right], \label{eq:FnslrG1} 
\eea
where
\bea
\ell_{\gamma} &=& \frac{p_0^{\gamma}}{F} + \frac{\langle x_0, p_0\rangle}{\left(1-|x_0|^2\right) K} x_0^{\gamma}, \\
\phi_{\mu} &=& \frac{2\kappa K}{(p_0^0)^3} (p_0^0\ell_{\mu}-F\delta_{0\mu}), \\
\phi_{\mu\nu} &=& \frac{2\kappa }{(p_0^0)^2} g_{\mu\nu}(x,p)-\frac{4\kappa K}{(p_0^0)^3}(\ell_\nu \delta_{0\mu} + \ell_\mu \delta_{0\nu}) + \frac{6\kappa K^2 }{(p_0^0)^4} \delta_{0\nu}\delta_{0\mu}.
\eea

Because $\phi(p_0)$ depends solely on $p_0$ rather than $x_0$, it follows that the second line in Eq. (\ref{eq:FnslrG1}) should not {it ad hoc} vanish. Due to the intricate nature of the geometric structure and the presence of additional curvatures as expressed in the second line, coupled with the absence of a theoretical framework to fully address the entirety of expression (\ref{eq:FnslrG1}), one is compelled to accept this unavoidable approximation. Only through this approximation can we gain insight into the implications of the proposed geometric quantization on the Riemann manifold. However, once it becomes feasible to incorporate all terms of Eq. (\ref{eq:FnslrG1}), it becomes necessary to reexamine the resulting quantized four-dimensional Riemann metric, as described in Eq. (\ref{eq:gmunuQ2}).

\section{Geometric Quantization, Geodesic Equations and Affine Connections}
\label{sec:GGA}

In accordance with Appendix \ref{sec:tildegmunuApp}, the suggestion was made to utilize the quantized fundamental tensor at a given point $x$ as a comprehensive representation of the curved spacetime in Riemann geometry, encompassing all relevant details \cite{Tawfik:2023ugm,Tawfik:2023hdi,Tawfik:2023rrm,Tawfik:2023kxq}
\bea
\tilde{g}_{\alpha\beta} &=& C(x,p)\, g_{\alpha\beta}. \label{eq:tildegalphabeta1}
\eea
The determination of the exact expression for the conformal coefficient $C(x,p)$ continues to pose a significant mathematical challenge. As an alternative, an approximate formulation has been put forward in the references \cite{Tawfik:2023ugm,Tawfik:2023hdi,Tawfik:2023rrm,Tawfik:2023kxq}
\bea
C(x,p) &=& \left(\phi(p_0)^{2}+2\frac{\kappa}{(p_{0}^0)^{2}} F^{2}\right)\left[1+\frac{\bar{m}^{2}}{{\mathscr F}^{2}} \left(1+2 \beta p_{0}^{\rho}p_{0 \rho}\right)\right]. \label{eq:Ctildegalphabeta1}
\eea
In the preceding text, it was noted that an averaging technique was utilized on $C(x,p)$. While this approach may lead to a bias towards a specific instance within the continuous spectrum ranges of the quantum operators forming $C(x,p)$, the truncated $C(x,p)$ continues to exhibit its important quantum-mechanical features.

The assumption of linearity in Eq. (\ref{eq:tildegalphabeta1}) implies that the relationship $-c^2 d\tau^{2} = ds^{2}=g_{\alpha\beta} dx^{\alpha} dx^{\beta}$ remains applicable to the quantized fundamental tensor $\tilde{g}_{\alpha\beta}$. By an appropriately parameterization, the proper time can be represented - in natural units - as $- d\tau^{2} = ds^{2}$
\bea    
\tilde{\tau}_{ab} &=& \int_{0}^{1}\sqrt{-\tilde{g}_{\alpha\beta}(x) \frac{dx^{\alpha}}{d\sigma}\frac{dx^{\beta}}{d\sigma}} = \int_{0}^{1}L\left(\frac{dx^{\alpha}}{d\sigma},x^{\alpha}\right)d\sigma.
\eea
By employing variational methods, similar to those used in classical dynamics, one can derive the Euler-Lagrange equations 
\bea
-\frac{d}{d\sigma} \frac{\partial L}{\partial (dx^{\gamma}/d\sigma)}+\frac{\partial L}{\partial x^{\gamma}}&=&0, \label{eq:L}
\eea
where $L$ is the Lagrangian and
\bea
\frac{\partial L}{\partial x^{\gamma}} &=& \frac{-L}{2}\left\{
C(x,p) \frac{\partial g_{\alpha \beta}}{\partial x^{\gamma}}\frac{d x^{\alpha}}{d\tau}\frac{d x^{\beta}}{d\tau} \right. \nn \\
&&\hspace*{5mm}\left. + g_{\alpha \beta}\frac{2\kappa}{(p_{0}^0)^{2}}\left[
1+\frac{\tilde{m}^{2}}{{\mathscr F}^{2}}\left(1+2 \beta p_{0}^{\rho} p_{0}^{\rho}\right)\right]
F^{2}_{,\gamma} \frac{d x^{\alpha}}{d\tau}\frac{d x^{\beta}}{d\tau}\right\}, \label{eq:L1} \hspace*{7mm} \\
\frac{d}{d\sigma}\frac{\partial L}{\partial (d x^{\gamma}/d\sigma)} &=& -L\left[\tilde{g}_{\alpha\gamma}\frac{d^{2} x^{\alpha}}{d\tau^{2}}+\frac{1}{2}\left(\frac{\partial \tilde{g}_{\alpha\gamma} }{\partial x^{\beta}}+\frac{\partial \tilde{g}_{\gamma\beta} }{\partial x^{\alpha}}\right)\frac{d x^{\alpha}}{d\tau}\frac{d x^{\beta}}{d\tau}\right]. \label{eq:L2}
\eea
The quantized geodesic equations are then derived by substituting Eqs. (\ref{eq:L1}) and (\ref{eq:L2}) into Eq. (\ref{eq:L})
\bea
\frac{d^{2}x^{\alpha}}{d\tau^{2}}+\tilde{\Gamma}^{\alpha}_{\delta \beta} \frac{dx^{\delta}}{d\tau}\frac{dx^{\beta}}{d\tau} + \frac{g_{\delta \beta}}{2 \tilde{g}_{\alpha \gamma}} K^{2}_{,\gamma} C(x,p) \frac{dx^{\delta}}{d\tau}\frac{dx^{\beta}}{d\tau} &=& 0. \label{eq:C1}
\eea
The geodesic equations obtained from the conventional fundamental metric tensor $g_{\alpha\beta}$ 
\bea
\frac{d^{2}x^{\alpha}}{d\tau^{2}}+\Gamma^{\alpha}_{\delta \beta} \frac{dx^{\delta}}{d\tau}\frac{dx^{\beta}}{d\tau}&=& 0, \label{eq:ClsGeoEq0}
\eea
and those obtained from $\tilde{g}_{\alpha\beta}$, as expressed in Eq. (\ref{eq:C1}), exhibit a remarkable difference. Such a comprehensive difference encompasses not only the entire third term in Eq. (\ref{eq:C1}), but also the additional quantum-mechanical ingredients and curvatures that arise from the quantized affine connections $\tilde{\Gamma}^{\alpha}_{\delta \beta}$ \cite{Tawfik:2023ugm,Tawfik:2023hdi,Tawfik:2023rrm,Tawfik:2023kxq}.
\bea
\tilde{\Gamma}^{\alpha}_{\delta \beta}= 
\Gamma^{\alpha}_{\delta \beta}+\frac{K^{2}_{, \gamma}}{2 C(x,p)}\left(\delta^{\alpha}_{\beta}+\delta^{\alpha}_{\delta}-g^{\alpha \gamma} g_{\delta \beta}\right). \label{eq:C}
\eea
Consequently, Eq. (\ref{eq:C1}) can be reformulated as,
\bea
\frac{d^{2}x^{\alpha}}{d\tau^{2}}+\Gamma^{\alpha}_{\delta \beta} \frac{dx^{\delta}}{d\tau}\frac{dx^{\beta}}{d\tau} + \frac{K^{2}_{, \gamma}}{2 C(x,p)}\left[\frac{g_{\delta \beta}}{\tilde{g}_{\alpha \gamma}} C^2(x,p) \frac{dx^{\delta}}{d\tau}\frac{dx^{\beta}}{d\tau} + \delta^{\alpha}_{\beta}+\delta^{\alpha}_{\delta}-g^{\alpha \gamma} g_{\delta \beta}\right] &=& 0. \label{eq:C2}
\eea
The third term represents the entire contributions that have arisen from the proposed geometric quantization approach.

\section{Evolution of Raychaudhuri Equations along Geodesic Congruence}
\label{sec:Raychau1}

For a set of kinematic variables including the quadratic invariant expansion or volume scalar $\Theta$, shearing $\sigma$, rotation $\omega$, and Ricci identity $R_{\mu \nu} u^{\mu} u^{\nu}$, the Raychaudhuri equations express the evolution equation along a geodesic congruence \cite{Raychaudhuri1957,Choudhury2021}
\bea
\frac{d\Theta}{d\tau} &=& - \frac{1}{3} \Theta^2 -  \sigma^2 + \omega^2 - R_{\mu \nu} u^{\mu} u^{\nu}, \label{eq:dThetaRaych1}
\eea
where $\sigma^2=\sigma_{\mu \nu} \sigma^{\mu \nu}$ and $\omega^2=\omega_{\mu \nu} \omega^{\mu \nu}$. In the evolution equation of the expansion, Ricci curvature tensor, $R_{\mu \nu}$ as well as Riemann and Weyl tensors appear along with the velocity field deriving the geodesic congruence. In Eq. (\ref{eq:dThetaRaych1}), the expansion tensor in dependence on $r$ and $t$ is given as
\bea
\Theta(r,t) &=& u^{\alpha}(r,t)_{;\alpha} = u^{\alpha}(r,t)_{,\alpha} + u^{\sigma}(r,t) \tilde{\Gamma}^{\alpha}_{\sigma \alpha} \nn \\
&=& \frac{1}{2 r a(t) \left(u^t(r,t)\right)^{-3}} \left[3 r \dot{a}(t) \left(u^t(r,t)\right)^{-2} - 6 a(t) \frac{\de M(r,t)}{\de t}
\right]. \label{eq:Thta}
\eea
The skew part of the velocities $u_{[\alpha ; \beta]}$ represents the rotation scalar
\bea
\omega_{\alpha \beta} &=& u_{[\alpha ; \beta]}. \label{eq:omega2a} 
\eea
With the conventional fundamental tensor $g_{\alpha \beta}$, the resulting finite component $\omega_{12} \omega^{12}$ leads to
\bea
\omega^2 &=& \left[\frac{100 a(t)}{r^2 \left[\Lambda + 8 \pi \tanh(t)\right]^2} - \frac{36 a(t)}{\Lambda + 8 \pi \tanh(t)} \right]^{-1}. \label{eq:omega2b} 
\eea 
The stress scalar reads
\bea
\sigma_{\alpha \beta} &=& u_{(\alpha ; \beta)} - \frac{1}{3} \Theta h_{\alpha \beta},  \label{eq:sigma2a} 
\eea
where $u_{(\alpha ; \beta)}$ are the symmetric parts of the velocities, $h_{\alpha \beta}=g_{\alpha \beta} + u_{\alpha} u_{\beta}$ is the spacial projection of the metric in directions orthogonal to the timelike vector fields $u_{\alpha}$, where $u_{\alpha} u^{\beta}=-1$ is the four-velocity normalization condition. 
%$\sigma^{11} \sigma_{11}$, $\sigma^{12} \sigma_{12}$, $\sigma^{22} \sigma_{22}$,  $\sigma^{33} \sigma_{33}$, and $\sigma^{44} \sigma_{44}$,
Then, the resulting nonvanishing components of $\sigma_{\alpha \beta}$ are combined  as
\bea
\sigma^{2} &=& \frac{\left(u^t(r,t)\right)^{6}}{r^2} \left[\frac{\de M(r,t)}{\de t}\right]^2 %\nn \\
%
%&+& 
+ \frac{\left(u^t(r,t)\right)^{6}}{r^2 a^2(t)} \left[r \left(u^t(r,t)\right)^{-2} \dot{a}(t) - a(t) \frac{\de M(r,t)}{\de t}\right]^2 \nn \\ 
&+& \frac{\left(u^t(r,t)\right)^{6}}{12 r^2 a^4(t)} \left[r\left[a(t)-1\right] \left(u^t(r,t)\right)^{-2} \dot{a}(t) + 2 \left[1-3a(t)\right] a(t) \frac{\de M(r,t)}{\de t} \right] \nn \\
&& \hspace*{12.5mm} \times \left[3 r \left[a(t)-1\right] \left(u^t(r,t)\right)^{-2} \dot{a}(t) + 6 \left[1-3 a(t)\right] a(t) \frac{\de M(r,t)}{\de t} \right] \nn \\
&+& \frac{\left(u^t(r,t)\right)^{6}}{9 r^2 a^2(t)}  \left[r \left(u^t(r,t)\right)^{-2} \dot{a}(t) + \left(-12 \pi r^3 \sech^2(t)+7 \frac{\de M(r,t)}{\de t} \right)a(t)\right]^2 \nn \\
&-& \frac{\left(u^t(r,t)\right)^{2}}{4 r^2 a(t)} \left[-4\pi r^2 \tanh(t) - \frac{\de M(r,t)}{\de r}\right]^2.
 \label{eq:sigma2b}
\eea
Now, we derive the fourth kinematic quantity,
\bea
R_{t t} u^{t} u^{t} &=& \left(u^t(r,t)\right)^2 \left\{\frac{3}{4}H^2(t) + 3 \left(u^t(r,t)\right)^{4} \left[-12 \left(\frac{\de M(r,t)}{\de t}\right)^2 + \left[-3 r + \Lambda r^3 + 6 M(r,t)\right] \frac{\de^2 M(r,t)}{\de t^2}\right] \right. \nn \\
&+&\left. \frac{\left(u^t(r,t)\right)^{2}}{6 r^2 a(t)}\left[
72 M^2(r,t)\left(\Lambda r + \frac{\de^2 M(r,t)}{\de r^2}\right) \right.\right. \nn \\
&& \left.\left. + 6 r M(r,t) \left(-9 r \ddot{a}(t) + 4 \left(-3+\Lambda r^2\right)\left(\Lambda r + \frac{\de^2 M(r,t)}{\de r^2}\right)\right) \right.\right. \nn \\
&& \left.\left.  + r^2 \left(-9r\left(-3+\Lambda r^2\right)\ddot{a}(t) + 45 \dot{a}(t) \frac{\de M(r,t)}{\de t} + 2 \left(-3+\Lambda r^2\right)^2 \left(\Lambda r + \frac{\de^2 M(r,t)}{\de r^2}\right)\right)
\right] \right\}. \hspace*{7mm} 
\label{eq:R11u11}
\eea

Then by substituting Eqs. (\ref{eq:Thta}), (\ref{eq:omega2b}), (\ref{eq:sigma2b}), and (\ref{eq:R11u11}) into Eq. (\ref{eq:dThetaRaych1}), we obtain the evolution of Raychaudhuri equations along the geodesic congruence expansion. The resulting analytic expression is numerically solved and was presented in Fig. \ref{fig:3}.

\section{Details Results with EGS Metric}
\label{App:EGA-Dtls}

%%%%%%%%%%%%%%%%%%%%%%%%%%%%%%%%%%%%%%
\begin{figure}[htb!] 
\includegraphics[angle=-90,width=0.6\textwidth]{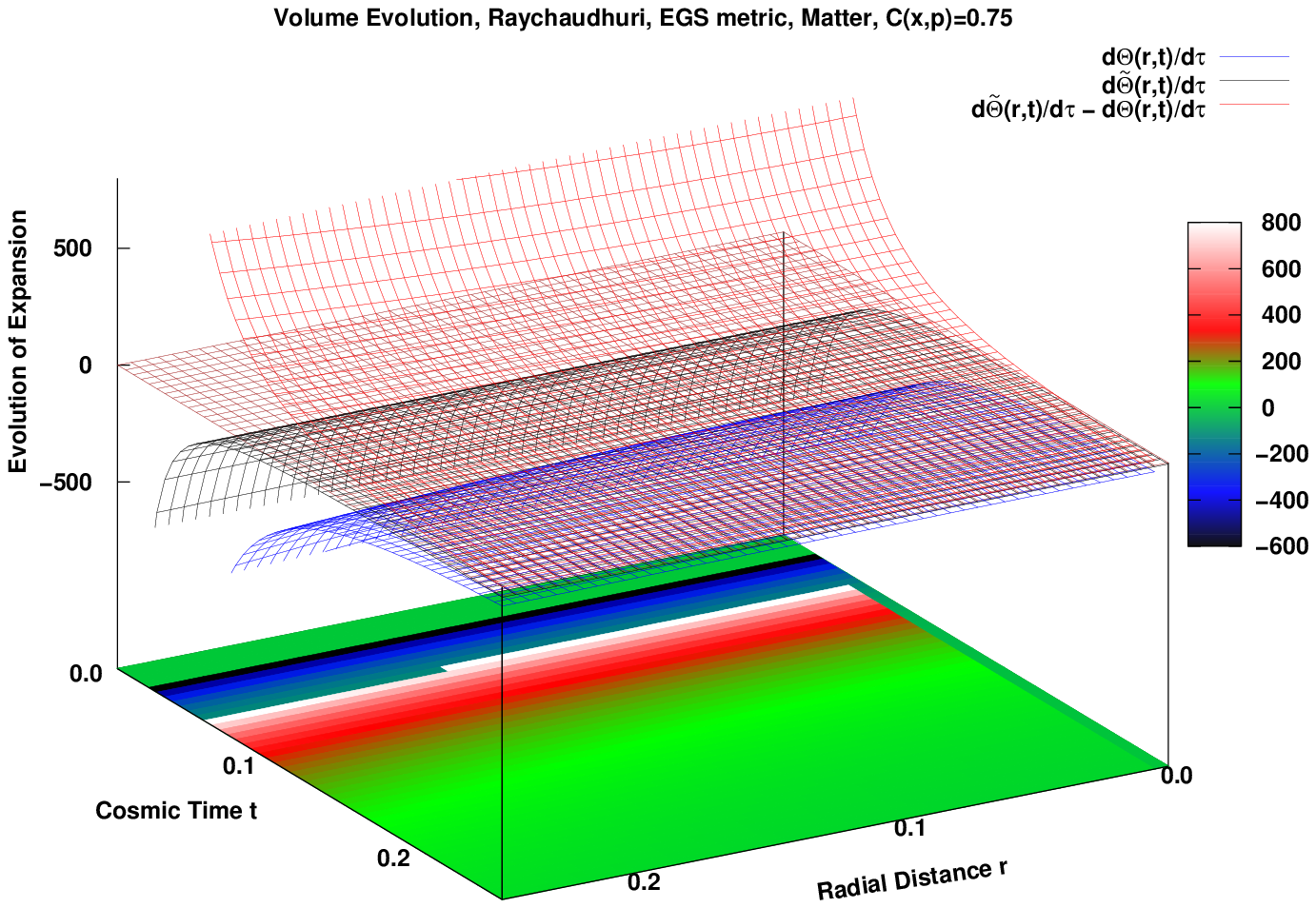}  
\includegraphics[angle=-90,width=0.6\textwidth]{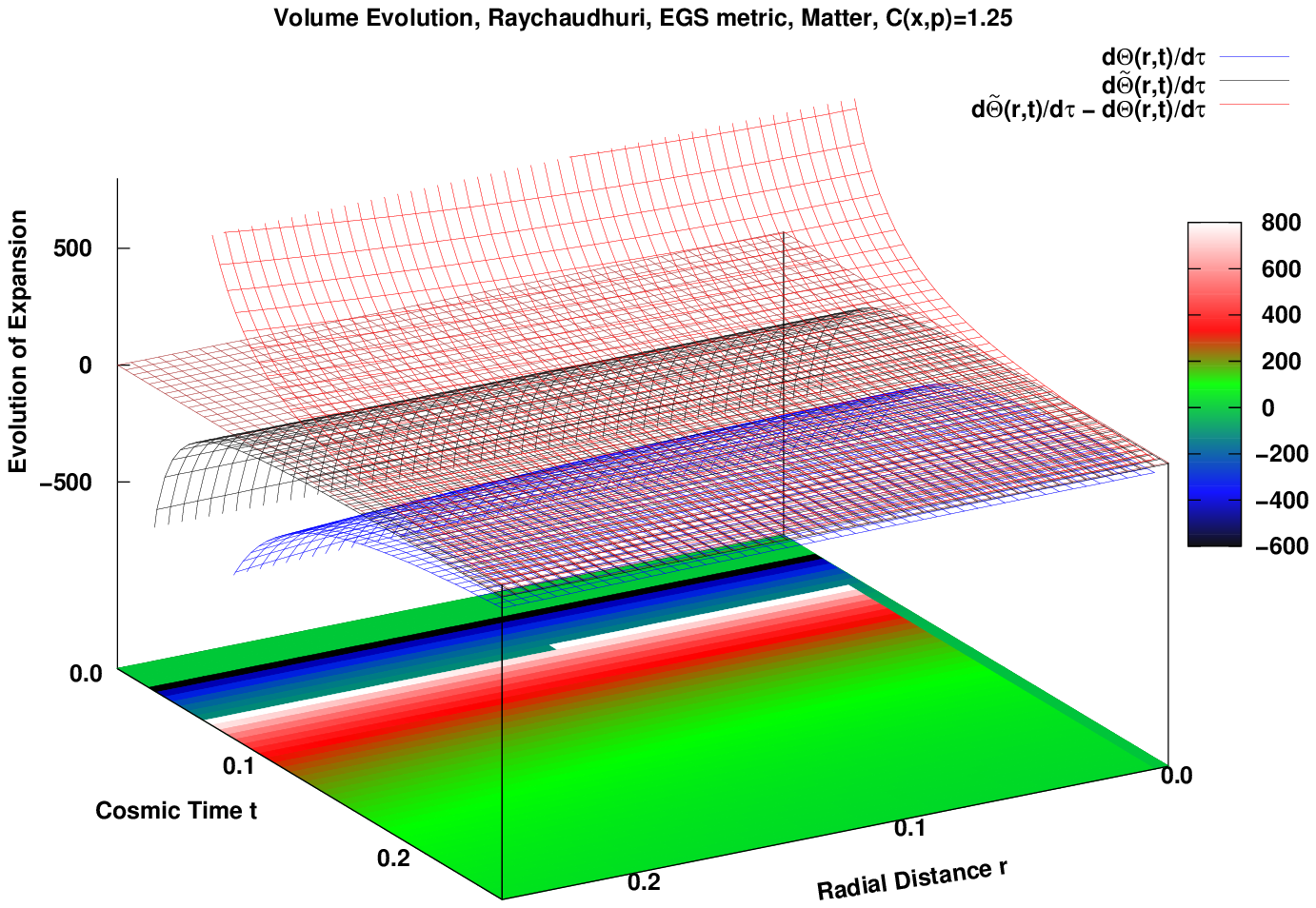} 
\caption{The numerical analysis of the volume scalar or expansion within the framework of the EGS metric is presented as a function of radial distance $r$ and cosmic time $t$. A systematic comparison is conducted among classical (blue lattice), quantum (red lattice), and hybrid (black lattice) results. For reference purposes, the results that describe the evolution of straight world-lines, which exhibit no curvatures, are illustrated as a gray lattice. \label{fig:3a}}
\end{figure}
%%%%%%%%%%%%%%%%%%%%%%%%%%%%%%%%%%%%%%

\begin{figure}[htb!] 
\includegraphics[angle=-90,width=0.6\textwidth]{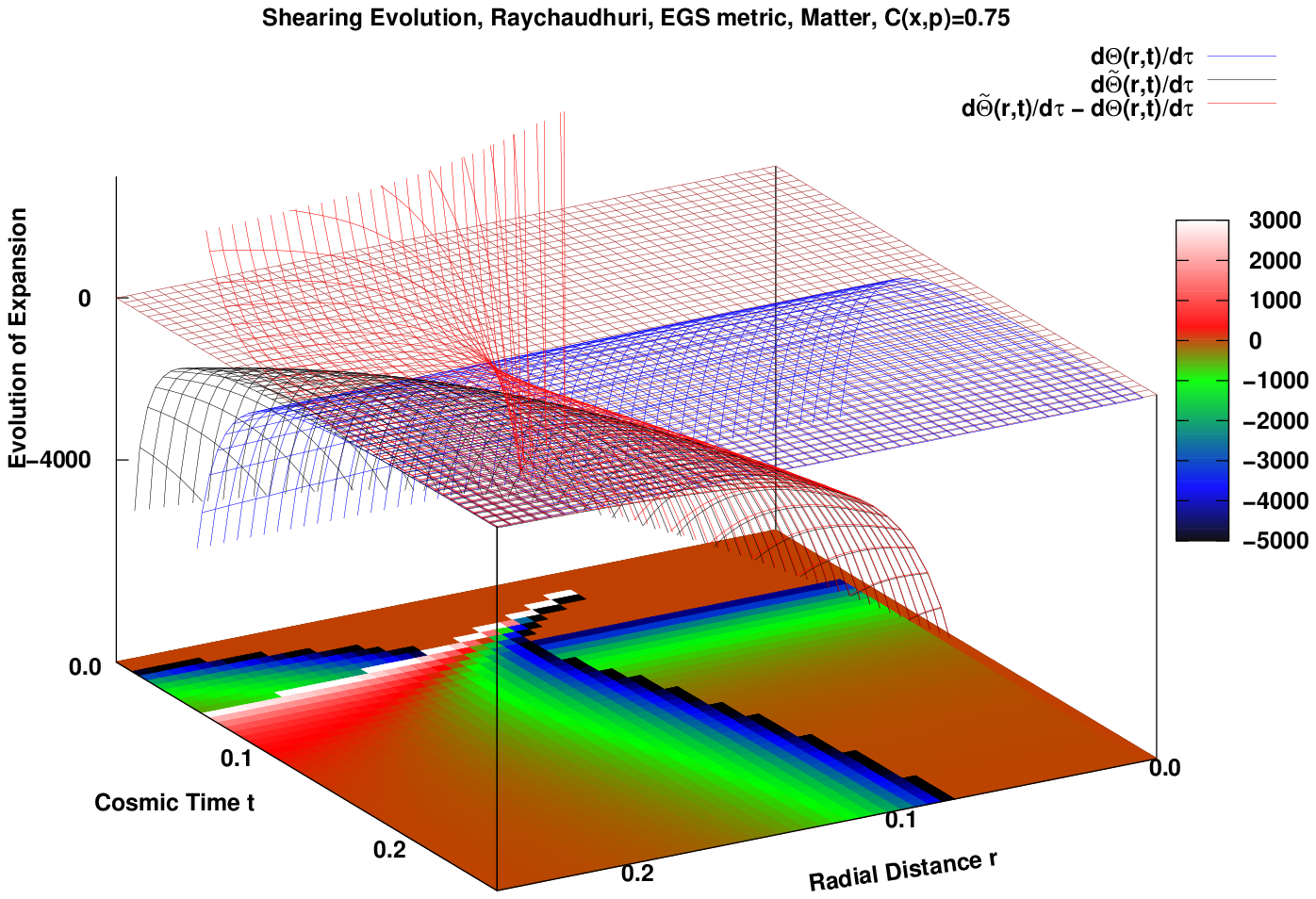}  
\includegraphics[angle=-90,width=0.6\textwidth]{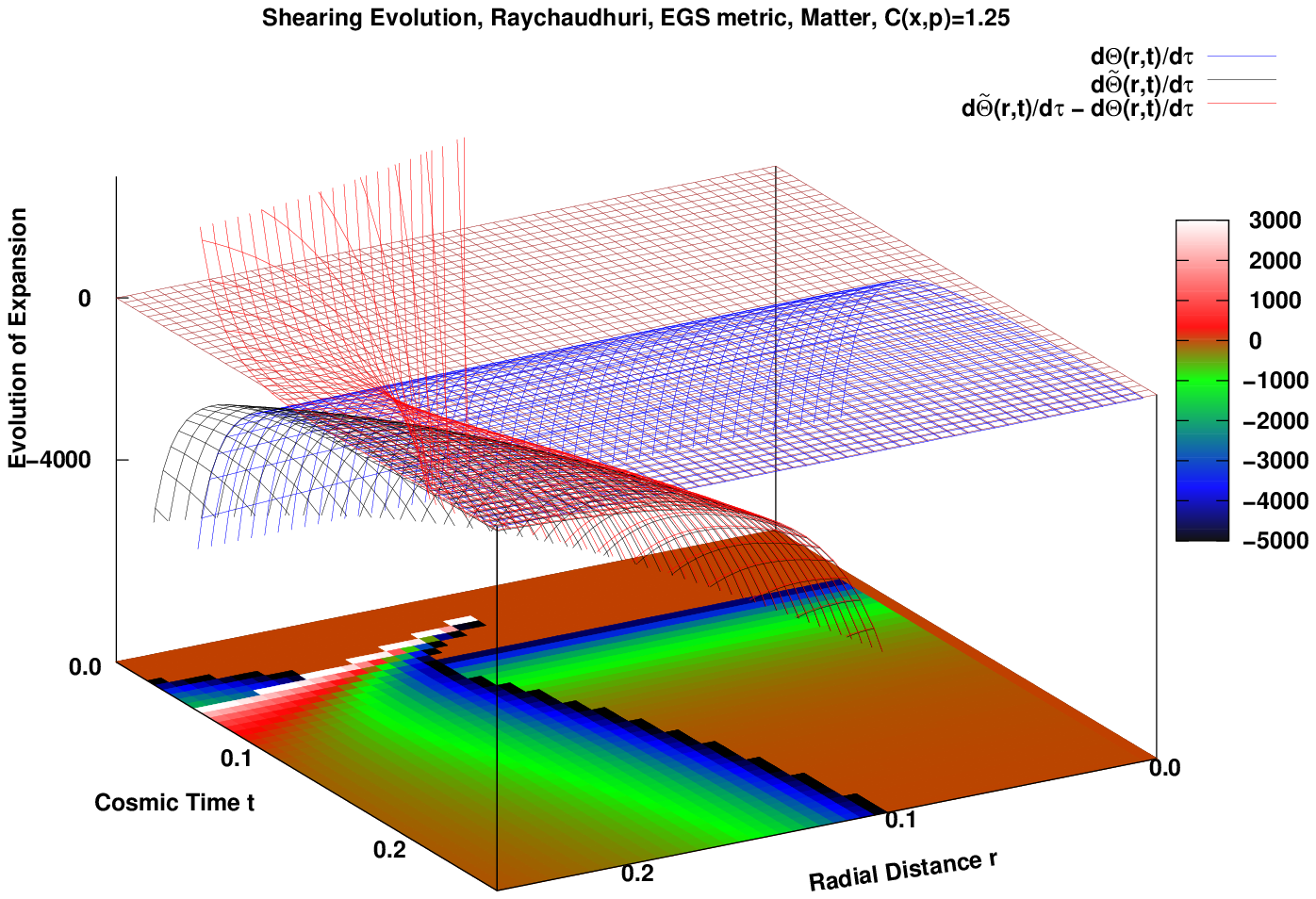} 
\caption{The same as in Fig. \ref{fig:3a} but here for shearing or anisotropy. \label{fig:3b}}
\end{figure}

\begin{figure}[htb!] 
\includegraphics[angle=-90,width=0.6\textwidth]{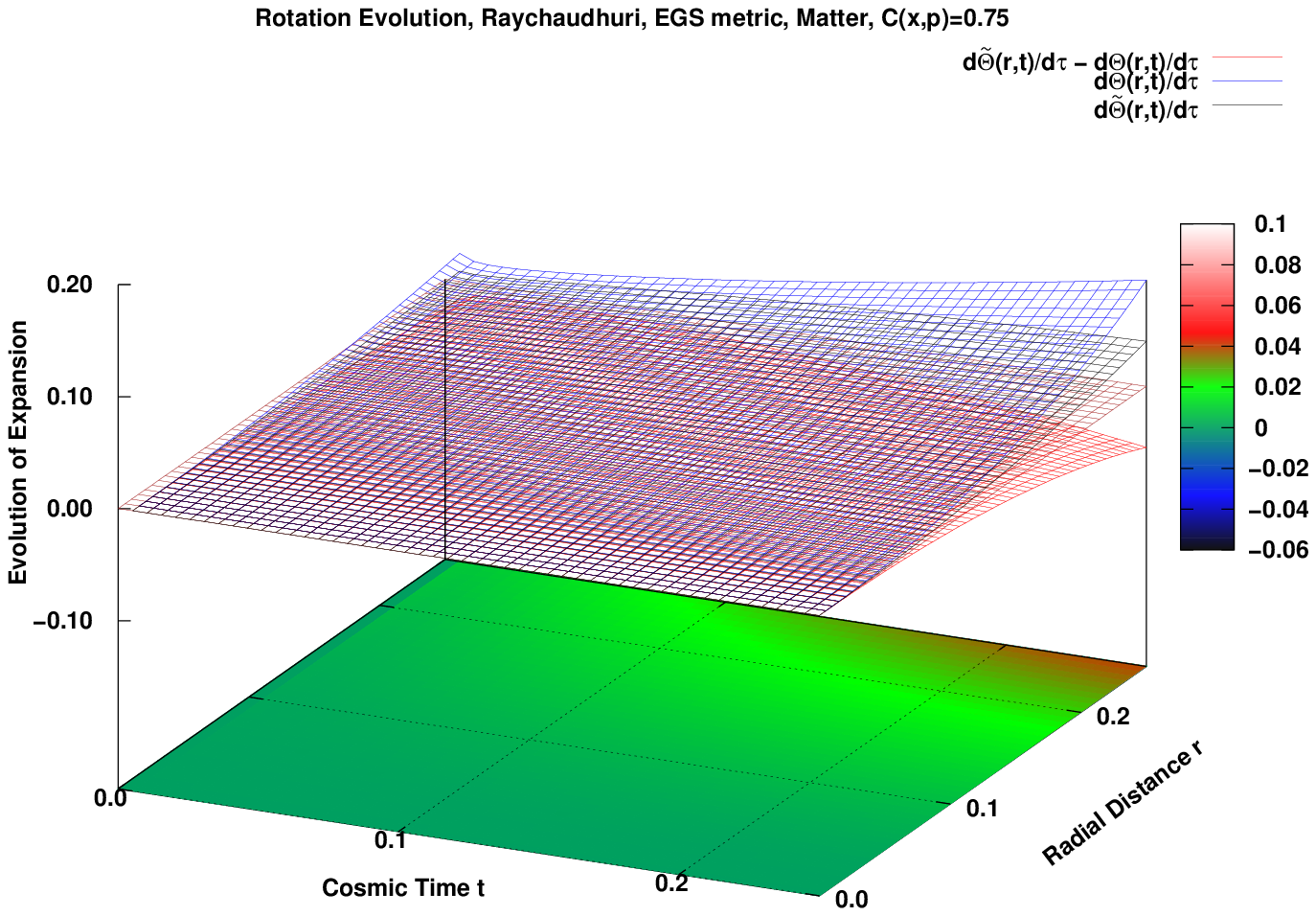}  
\includegraphics[angle=-90,width=0.6\textwidth]{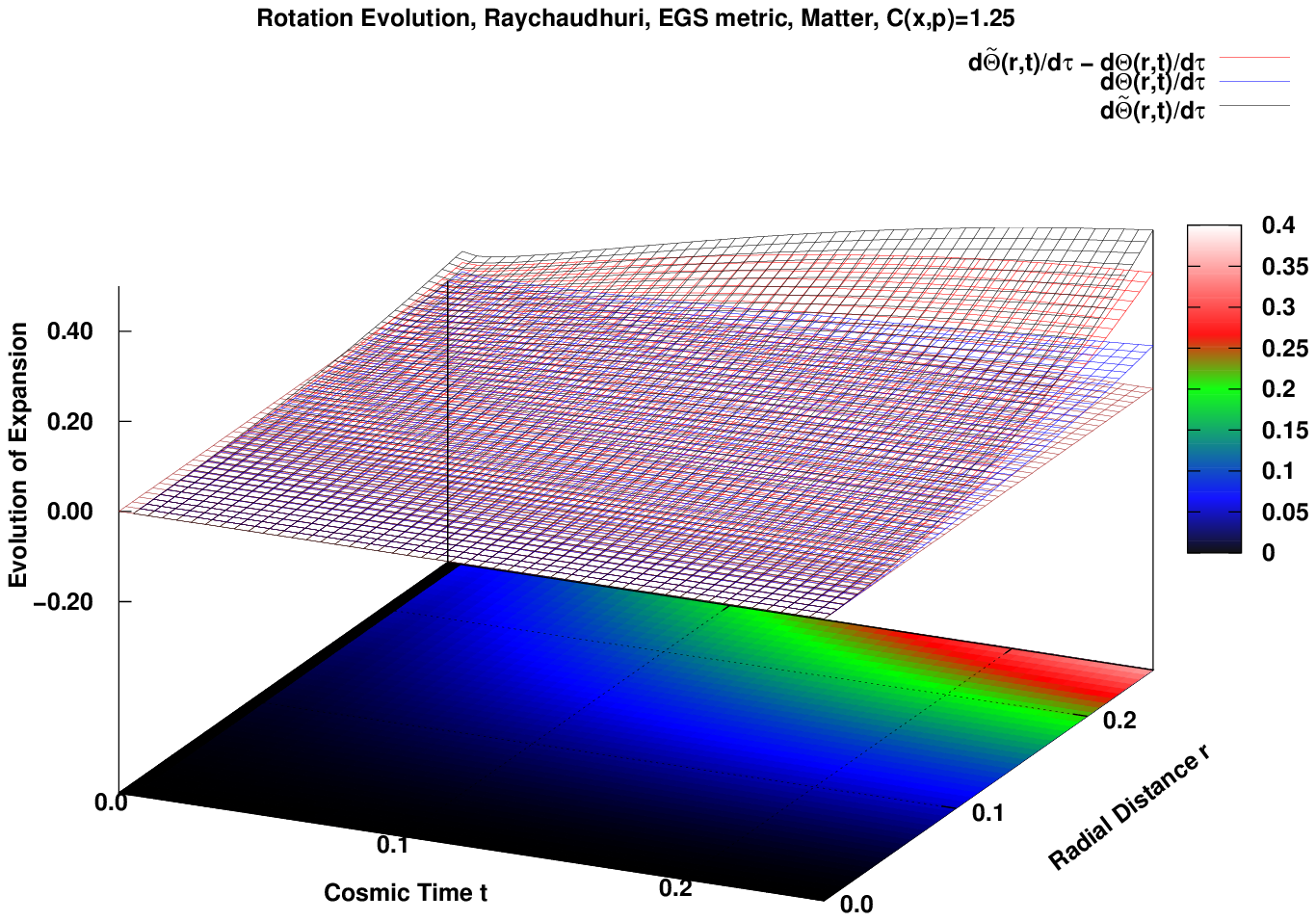} 
\caption{The same as in Fig. \ref{fig:3a} but here for rotation or spinning. \label{fig:3c}}
\end{figure}

\begin{figure}[htb!] 
\includegraphics[angle=-90,width=0.6\textwidth]{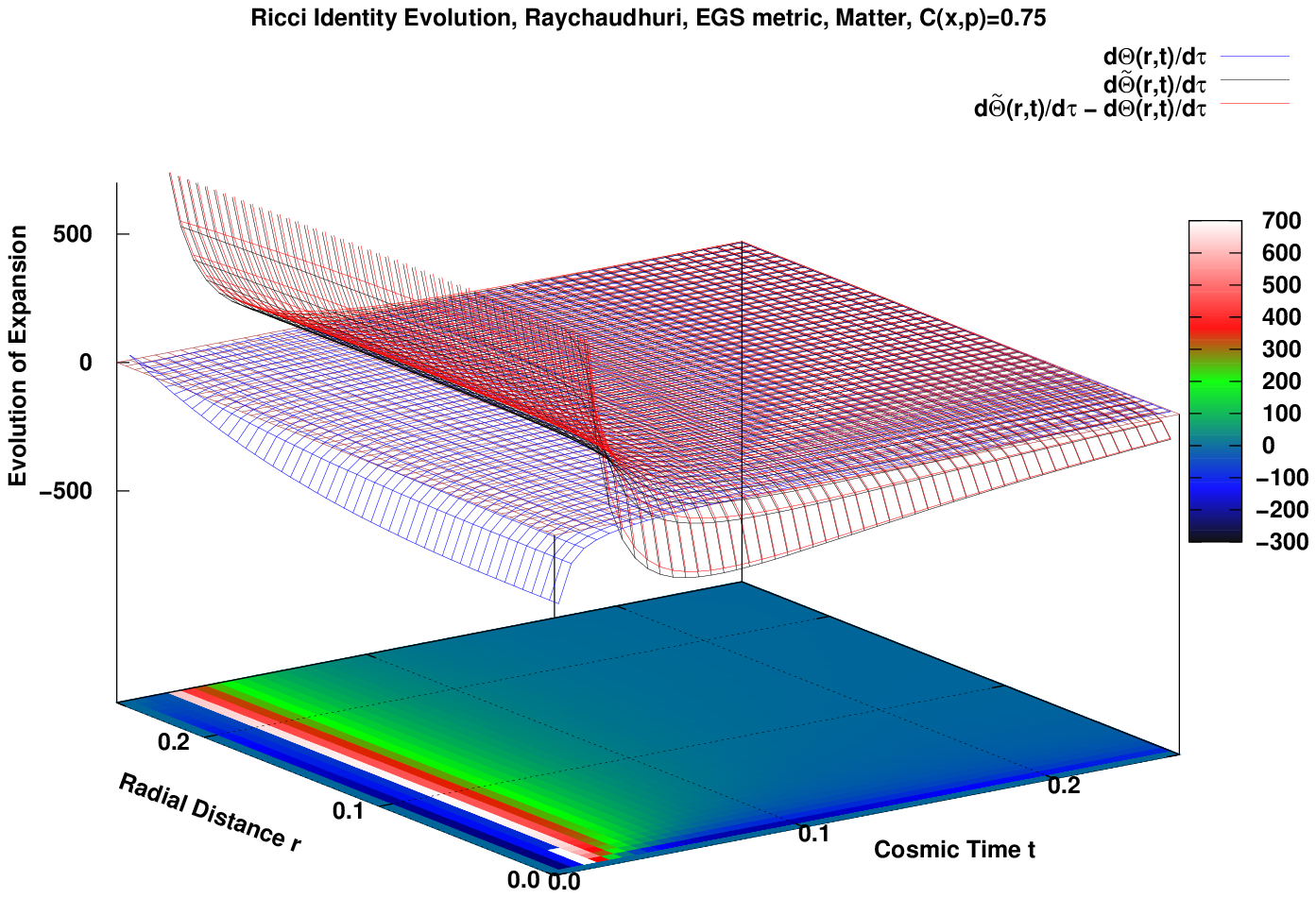}  
\includegraphics[angle=-90,width=0.6\textwidth]{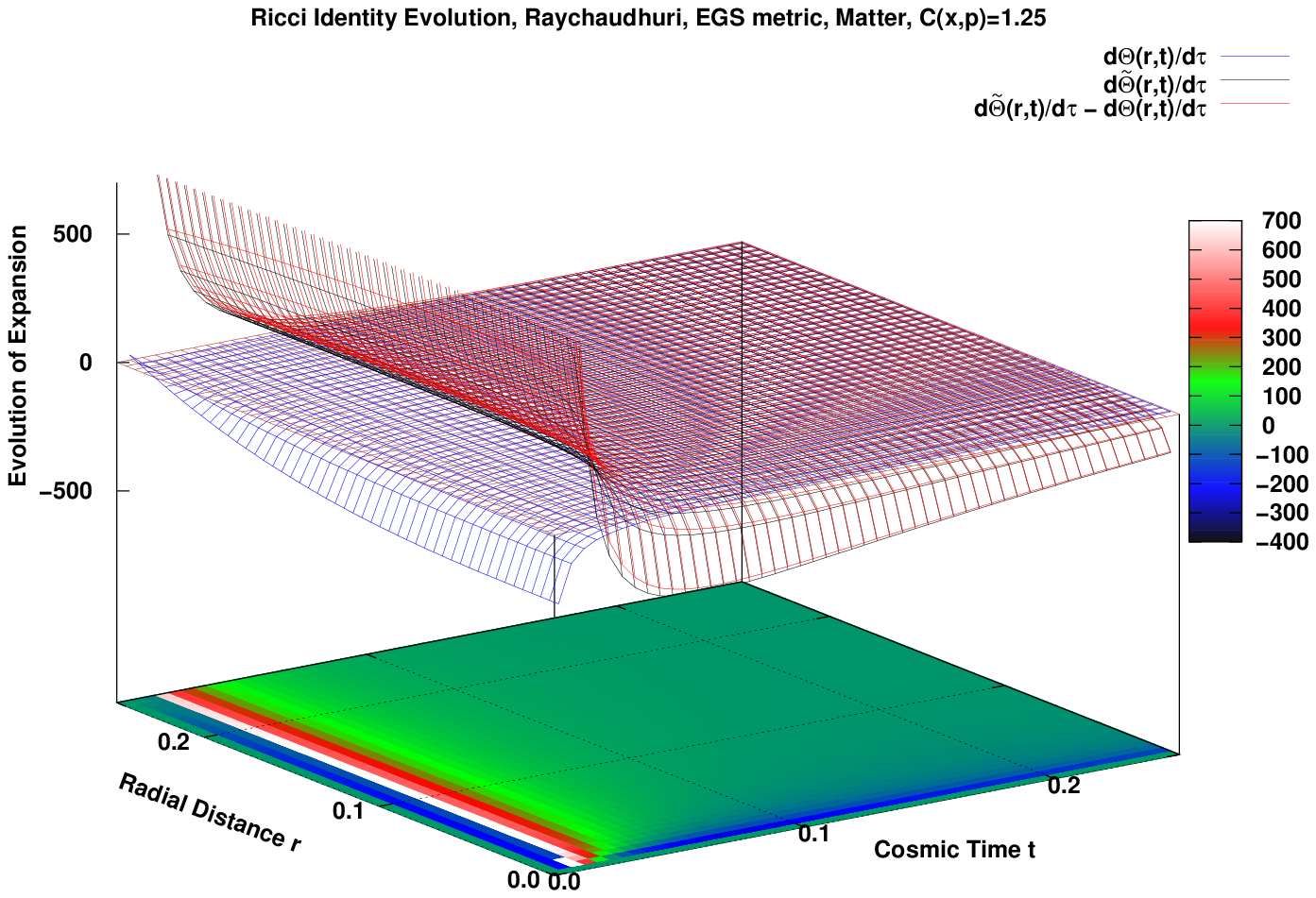} 
\caption{The same as in Fig. \ref{fig:3a} but here for Ricci identity or local gravitational field. \label{fig:3d}}
\end{figure}

The data presented in Figs. \ref{fig:3a}-\ref{fig:3d} represent the volume scalar (expansion), shearing (anisotropy), rotation (spinning), and Ricci identity (local gravitational field), respectively. Such an in-depth examination allows for the discernment of the dynamic quantities that play a role in the aggregate results depicted in Fig. \ref{fig:3}.

The volume scalar (expansion) depicted in Fig. \ref{fig:3a} is derived from the EGS solution of the Einstein Field Equations (EFE) and is expressed as a function of the parameters $r$ and $t$. As demonstrated in prior figures, a comparative analysis is conducted between the results associated with the conventional metric tensor (blue lattice) and those linked to the quantized metric tensor (black and red lattices). The analysis reveals that the conventional metric tensor is linked to a negative volume scalar (expansion), which experiences a substantial reduction with the decrease in cosmic time. The effect of radial distance is notably less significant.

The contributions from pure quantum contributions show an inverse correlation with the parameter $t$ (red lattice). As $t$ decreases, there is a marked increase in the volume scalar (expansion). All results are positive in all instances. The evaluation of outcomes in comparison to straight world-lines, defined by a lack of curvature and evolution, emphasizes the presence of both favorable and unfavorable progressions. However, the relationship with the variable $r$ remains minimal. The degree of quantization has a minimal impact, leading to results at $C(x,p)=0.75$ in the top panel that closely resemble those at $C(x,p)=1.25$ in the bottom panel.

%%%%%%%%%%%%%%%%%%%%%%%%%%%%%%%%%%%%%%
%%%%%%%%%%%%%%%%%%%%%%%%%%%%%%%%%%%%%%

The shearing (anisotropy) as a function of $r$ and $t$ is illustrated in Fig. \ref{fig:3b}, revealing a complex structure. The results derived from the conventional metric tensor are characterized by smoothness and continuity, as indicated by the blue lattice. This suggests that the shearing (anisotropy) remains consistently negative and decreases with decreasing $t$, exhibiting no significant dependence on $r$. Conversely, the results obtained from the quantized metric tensor, depicted by the red lattice, reveal a notable correlation with both $r$ and $t$. At higher values of $r$ and $t$, the shearing (anisotropy) aligns closely with the classical metric tensor results (blue lattice) and the evolution of flatness (gray lattice). The simultaneous decrease of $r$ and $t$ results in a significant transformation in the structure of shearing (anisotropy). The dependence observed is characterized by a lack of smoothness and continuity. When $r$ is small, the shearing (anisotropy) exhibits a rapid increase in the positive direction, while small values of $t$ are associated with a swift increase in the positive direction. This sudden transition between positive and negative shearing (anisotropy) underscores the quantum contributions, leading to a radical reinterpretation of classical results in the context of the hybrid results (black lattice). Consequently, the shearing (anisotropy) experiences a significant decline as both $r$ and $t$ are decreased. It is also noteworthy that the degree of quantization plays a crucial role in influencing shearing, thereby demonstrating anisotropic properties.

%%%%%%%%%%%%%%%%%%%%%%%%%%%%%%%%%%%%%%
%%%%%%%%%%%%%%%%%%%%%%%%%%%%%%%%%%%%%%

Figure \ref{fig:3c} illustrates the relationship between rotation (spinning) and the variables $r$ and $t$. It is observed that decreasing both $r$ and $t$ leads to an slight increase in the classical and hybrid results. Conversely, the rotation (spinning) derived from the quantum metric tensor exhibits a slight decrease, particularly at $c(x,p)=0.75$. However, at $c(x,p)=1.25$, decreasing $r$ and $t$ results in an increase in quantum rotation (spinning). Notably, the magnitudes of these results are insufficient to significantly influence the evolution of the Raychaudhuri equations.

In relation to the results obtained from $g_{\alpha \beta}$, it is essential to remember that the idea of vanishing rotation is conjectured to be connected to hypersurface orthogonal vector fields. On the other hand, finite shearing is tied to the Frobenius theorem, a classical theorem in differential geometry that correlates vector fields with sub-manifolds of a smooth manifold. The interactions between rotation and shearing, along with their relationships on smooth Riemannian manifolds, have been discussed in the context of the Frobenius theorem \cite{Eric:2004ngn}.

%%%%%%%%%%%%%%%%%%%%%%%%%%%%%%%%%%%%%%
%%%%%%%%%%%%%%%%%%%%%%%%%%%%%%%%%%%%%%

The local gravitational field, represented by the Ricci identity, is depicted in Fig. \ref{fig:3d}. Our results indicate that the conventional metric tensor yields results that are negligible in most cases, except at very low values of $r$ and $t$. The decrease observed at small $t$ is significantly more substantial than that at small $r$. The degree of quantization is nearly imperceptible. However, with quantization, the variation at small $r$ and $t$ is radically reversed, resulting in a rapid increase at small $t$, which is associated with a swift decrease at small $r$. This quantization approach, similar to the numerical analysis of shearing or anisotropy, reveals the true double dependence of the EGS metric on both $r$ and $t$.  

In conclusion, the variations among the various kinematic quantities serve as the basis for the preference towards either a non-singular or singular Universe. It is evident that the terms on the right-hand side of Eq. (\ref{eq:dThetaRaych2}) that support recollapse - namely, the finite volume scalar and shearing - dominate those that oppose it, such as finite rotation or vorticity. The primary contributions to this outcome stem from the volume scalar and shearing effects. While the influence of vorticity is nearly negligible, the local gravitational field provides minor positive contributions, particularly at smaller values of $r$ and $t$. Additionally, there are two terms that are omitted from Eq. (\ref{eq:dThetaRaych2}). The first term is the strong energy condition, represented by the tidal tensor, which has a positive trace that promotes recollapse. The second term gives the acceleration fields arising from self-gravitation, which, due to their positive divergence, act against recollapse.

\section*{Acknowledgment}

AT and MN would like to recognize the financial backing received from the Egyptian Academy for Scientific Research and Technology, provided through the ASRT/JINR joint projects (Call 2023), which has made it possible to conduct a segment of this research at the Joint Institute for Nuclear Research (JINR) in Russia.

\section*{Conflicts of Interest}

The authors declare that there are no conflicts of interest regarding the publication of this published article!

%The authors declare that the present script is in compliance with ethical standards regarding its content!

\section*{Dataset Availability}

The data used to support the findings of this study are included within the published article and properly cited! All of the material is owned by the authors.
%
%All data generated or analyzed during this study are included in this published article. 

%\section*{Author contributions}
%
%The responsibility for proposing the conception of the present study lies with AT, who also undertook the tasks of designing and managing the research, interpreting the results, deriving the expressions, drawing the figures, and preparing the manuscript. AA contributed to the writing and proofreading of the manuscript. The final version of the manuscript was unanimously approved by all authors.

%\section*{Funding}
%
%The authors declare that this research received no specific grants from any funding agency in the public, commercial, or not-for-profit sectors.

%The funding agencies have no role in the design of the study; in the collection, analysis or interpretation of the data; in the writing of the manuscript; or in the decision to publish the results!

\section*{Competing interests}

The authors confirm that there are no relevant financial or non-financial competing interests to report.

\bibliographystyle{unsrtnat}
\bibliography{ListOfReferences-Motion}

\end{document}